\pdfoutput=1
\documentclass[useAMS,usenatbib]{mn2e}
\usepackage{graphicx}
\usepackage{epsfig}
\usepackage{amsmath}
\usepackage{pdfpages}
\usepackage[T1]{fontenc}
\newcommand{\beq}{\begin{equation}}
\newcommand{\eeq}{\end{equation}}
\newcommand{\beqa}{\begin{eqnarray}}
\newcommand{\eeqa}{\end{eqnarray}}

\def\deg{\,{\rm deg}}
\def\km{\, \textrm{km}}
\def\s{\, \textrm{s}}
\def\Mpc{\, \textrm{Mpc}}

\def\apj{ApJ}

\def\apjs{ApJS}
\def\mnras{MNRAS}
\def\aap{A\&A}
\def\physrep{Phys.Rev}
\def\prd{Phys.Rev.D}

\def\aj{AJ}
\def\logm{\log(M_\ast/M_\odot)}

\begin{document}

\title{Probing the galaxy-halo connection in UltraVISTA to $z\sim2$}

\author[H.J. McCracken et al.] {H.~J. McCracken$^1$\thanks{E-mail: hjmcc@iap.fr (HJMCC)},
 M.~Wolk$^{1,2}$, S.~Colombi$^1$, M.~Kilbinger$^{1,3}$, O.~Ilbert$^4$,
\newauthor
 S.~Peirani$^1$, J.~Coupon$^5$, J.~Dunlop$^6$, B.~Milvang-Jensen$^7$, K. Caputi$^8$,
\newauthor
H. Aussel$^3$, M. B\'ethermin$^9$ and O. Le F\`evre$^4$.  
\\
(Affiliations after references)
}
\maketitle
\date{\fbox{\sc Draft Version: \today}}

\begin{abstract} \\

  We use percent-level precision photometric redshifts in the
  UltraVISTA-DR1 near-infrared survey to investigate the changing
  relationship be- tween galaxy stellar mass and the dark matter
  haloes hosting them to $z\sim2$. We achieve this by measuring the
  clustering properties and abundances of a series of volume-limited
  galaxy samples selected by stellar mass and star-formation
  activity. We interpret these results in the framework of a
  phenomenological halo model and numerical simulations. Our
  measurements span a uniquely large range in stellar mass and
  redshift and reach below the characteristic stellar mass to
  $z\sim2$. Our results are: 1. At fixed redshift and scale,
  clustering amplitude depends monotonically on sample stellar mass
  threshold; 2. At fixed angular scale, the projected clustering
  amplitude decreases with redshift but the co-moving correlation
  length remains constant; 3. Characteristic halo masses and galaxy
  bias increase with increasing median stellar mass of the sample;
  4. The slope of these relationships is modified in lower mass
  haloes; 5. Concerning the passive galaxy population, characteristic
  halo masses are consistent with a simply less-abundant version of
  the full galaxy sample, but at lower redshifts the fraction of
  satellite galaxies in the passive population is very different from
  the full galaxy sample; 6. Finally we find that the ratio between
  the characteristic halo mass and median stellar mass at each
  redshift bin reaches a peak at ${\log(M_h/M_\odot)}\sim12.2$ and the
  position of this peak remains constant out to $z\sim2$. The
  behaviour of the full and passively evolving galaxy samples can be
  understood qualitatively by considering the slow evolution of the
  characteristic stellar mass in the redshift range probed by
  our survey.

\end{abstract}


\section{Introduction}
\label{sec:introduction}

How are galaxies distributed in dark matter haloes? What is the
relationship between visible galaxies and the invisible dark matter?
How do the characteristics of these dark matter haloes control the
process of galaxy formation? In recent years, considering the dark
matter haloes hosting galaxies has provided an alternative perspective
on the galaxy formation question by permitting a consideration of how
halo mass can regulate star-formation activity.

Although galaxy-galaxy lensing, using the distortion of distant
background objects, can measure foreground dark matter mass
distributions \citep{Leauthaudetal10}, and can provide
direct information concerning dark matter halo masses and mass
profiles, this technique is challenging observationally and can only
probe a relatively narrow redshift baseline as the background galaxy
population must be resolved. Even with space-based observations, it is
challenging to apply this technique above $z>1$ as foreground galaxies
become unresolved.

A simpler although more indirect approach is to compare the observed
abundance and clustering properties of the galaxy samples with
predictions of phenomenological ``halo'' models
\citep{Scoccimarro:2001p11099,Seljak:2000p1153,Peacock:2000p11176,Neyman:1952p11742}. These
models contain an empirical prescription describing how galaxies
populate dark matter haloes (the ``halo occupation function'') : their
drawback is that they rely on a an accurate knowledge of the halo mass
function and halo profile, which must be calibrated using numerical
simulations. Although there is some doubt over the applicable regime
for these calibrations and the importance of second order effects such
as ``halo assembly bias'' \citep{Croton:2007p11858,Zentner:2014ki},
these techniques remain promising for high-redshift observations where
cosmic variances and sample errors are the most important source of
uncertainty and systematic errors do not yet dominate. (It is worth
mentioning that these assembly bias effects have been shown to be
important only in catalogues several orders of magnitudes larger this
present work.) A related technique involves comparing the abundances
of dark matter haloes with those of high-resolution N-body
simulations, the ``sub-halo abundance matching'' technique
\citep{2006ApJ...647..201C}.

The principal advantage of these methods is that they may be applied
over a relatively large redshift baseline and require as observations
only abundance and clustering measurements. A considerable industry
has developed in recent years
\citep{Behroozietal13,Mosteretal10,2009ApJ...696..620C} in applying
variants of this model in different redshift ranges and samples and
attempting to interpret these results in terms of models of galaxy
formation and evolution of star formation activity. However,
understanding the derived halo masses and halo occupation functions
over large redshift baselines--- at least to $z\sim2$ --- has been
complicated by the difficulty of comparing diverse data sets with
different selection functions. For example, not all samples are
cleanly selected in stellar mass and may use either luminosity or even
star-formation rate selection (as is the case with colour-colour
selected ``BzK'' \citep{Daddi:2004p76} galaxies. A final difficulty is
that until now most surveys are not deep enough to reach below the
all-important characteristic halo mass $M^*$ at least to $z\sim1$, and
for this reason have concentrated primarily on more massive galaxies
\citep{Foucaud:2010p11288}; those that \textit{are} deep enough have
instead been unable to constrain the more massive end because of
insufficient area \citep{Bielby:2014dv}. Reaching below this mass
limit is important to understand how the rapid build-up of the faint
end of the mass function occurs at $z\sim1$.

The COSMOS field \citep{Scoville:2007p12720} provides a bridge between
small, high redshift surveys (like GOODS, CANDELS and other HST deep
legacy fields) and larger, intermediate and local surveys like the
Canda-France Legacy Survey \citep{Couponetal12} and the SDSS
\cite{Zehavietal11}. One of its principal advantages is that it
contains a unique collection of spectroscopic redshifts fully sampling
$0.5<z<4$ and multi-band photometry \cite{2007ApJS..172...99C}
allowing a precise calibration of photometric redshifts, providing a
precision better than $1\%$ for both passive and star-forming
galaxies.  By using broad-band COSMOS data, in combination with the
DR1 \mbox{UltraVISTA} near-infrared $YJHK_{\rm~s}$ data
\citep{McCracken:2012gd}, we are able to accurately determine stellar
masses at least until $z\sim4$ for samples as faint as
$K_{\rm s}<24.0$ \citep{Ilbert:2013dq}. Although large local
spectroscopic redshift surveys such as BOSS and SDSS have
revolutionised our knowledge of the distribution of galaxies in the
local Universe, together with more recent magnitude-selected and
color-selected surveys such as VIPERS, VVDS and DEEP2
\citep{2003SPIE.4834..161D,Lefevretal05b,Guzzo:2014eb} which allowed
measurements to be extended to $z\sim1$ it is only photometric
redshift surveys which can probe the distribution of the galaxy
population over such a broad redshift baseline in a single sample to
relatively uniform limits. The unique aspect of this current work is
our precise stellar mass measurements over a large redshift range. 

Previous studies of the halo occupation distribution (the relationship
between the number of galaxies in a give dark matter halo and the halo
mass) have revealed that the evolution of the angular correlation
function and the distribution of satellite galaxies can be adequately
explained by an unchanging halo occupation distribution. Some evidence
has also emerged that the host halo mass at which the mass in stars
reaches a maximum moves slowly towards higher halo masses
\citep{Leauthaudetal12,Couponetal12}. In other words, over the
lifetime of a halo, star-formation processes occur more efficiently in
more massive haloes, and it is tempting to draw a link between this
relationship and the observed luminosity-dependant nature of galaxy
star-formation rate \citep{Cowie:1996p8471}.

Our aim in this work is to use the large, uniform, COSMOS-UltraVISTA
photometric redshift catalogue to investigate the relationship between
dark matter halo masses and the properties of the visible galaxy
population.  We will do this by comparing observations (clustering and
abundances) of a series of mass-selected galaxy samples to predictions
of a theoretical halo model and also to the results of the sub-halo
abundance matching using a high-resolution numerical simulation. 
In this paper we use a flat $\Lambda$CDM cosmology ($\Omega_{\rm
  m}=0.27$, $\Omega_{\Lambda}=0.73$, $H_{0}= 100 h \km \s^{-1}
\Mpc^{-1}$ and $\sigma_{8}=0.8$) with $h = 0.7$. All magnitudes are
in the ``AB'' system  \citep{Oke:1974p12716}.

\section{The UltraVISTA-COSMOS survey}

\subsection{Survey overview and photometric redshift estimation}
\label{sec:surv-overv-phot}

We use the publicly-available UltraVISTA-COSMOS photometric redshift
catalogue. This is a near-infrared selected galaxy sample, where
objects are detected on a very deep $YJHK_{\rm s}$ ``chisquared''
\citep{Szalay:1999p4804} detection image: the advantage compared to a
simple $K_{\rm s}$ catalogue that is many more bluer sources are
included. A complete description of the photometric redshifts derived
from this catalogue, and their parent photometric catalogue can be
found in \cite{Ilbert:2013dq}. The UltraVISTA DR1 release
\cite{McCracken:2012gd} covers $1.5~\mathrm{deg}^2$ of the COSMOS
field with deep $YJHK_{\rm s}$ data at least one or two magnitudes
deeper than the previous \citep{2010ApJ...708..202M} COSMOS
near-infrared data.

Photometric redshifts were derived using ``Le
Phare''\footnote{\texttt{http://www.cfht.hawaii.edu/arnouts/lephare.html}}
\citep{Ilbertetal06, Couponetal09}. ``Le Phare'' is a standard
template fitting procedure using 31 templates including elliptical and
spiral galaxies from the \cite{Pollettaetal07} library and 12
templates of young blue star-forming galaxies from
\cite{Bruzualetal03} stellar population synthesis models. Following
standard procedure, the templates are redshifted and integrated
through the instrumental transmission curves. The opacity of the
intra-galactic medium is accounted for and internal extinction can be
added as a free parameter to each galaxy. Photometric redshifts are
derived by comparing the modeled fluxes and the observed fluxes with a
$\chi^{2}$ merit function. In addition, a probability distribution
function is associated to each photometric redshift. A detailed
exploration of the precision of these photometric redshifts at
intermediate redshifts has been carried out in
\citeauthor{Ilbert:2013dq} using a unique, large sample of
spectroscopic redshifts covering the redshift range $1<z<3$.

We use the usual $\sigma_{\Delta_{z}}/(1+z_{s})$ estimator where
$z_{p}$ and $z_{s}$ are the photometric and the spectroscopic
redshifts respectively and $\Delta_{z} = z_{p} - z_{s}$.
``Catastrophic'' redshift errors are defined as objects with $|z_{p} -
z_{s}|/(1 + z_{s}) > 0.15$. The percentage of these objects is denoted
by $\eta$. Errors were estimated using the normalised median absolute
deviation: $1.48 \times$ median($|z_{p} - z_{s}|/(1 + z_{s})$)
\citep{Hoaglinetal83}.

For our $K_{\rm s}<24$ cut at $z < 1.5$ our photometric redshifts
have a precision of better than 1$\%$ and with less than 1$\%$ of
catastrophic failures. At $1.5 < z < 2$, the precision remains
excellent at $\sim0.03$ with the percentage of catastrophic failures
less than $\sim 1\%$. Even at $1.5 < z < 4$, the precision is
approximately 3$\%$ with around $7\%$ of catastrophic
failures. We limit to our analysis to $z<2.5$: above this redshift
range, the number of sources becomes too small to reliably measure
clustering or abundances and our cross-bin contamination becomes
significant as we will see in Section~\ref{sec:effect-phot-redsh},
where we investigate the effect of photometric errors on our
clustering measurements. 

We also use stellar masses computed in \cite{Ilbert:2013dq}. The
\cite{Bruzual:2003p963} stellar population synthesis models with a
\cite{Chabrier:2003ki} initial mass function is used to generate a
library of synthetic spectra normalised at one solar mass, which are
then fitted to the photometric measurements described above using ``Le
Phare''. As explained in \citeauthor{Ilbert:2013dq}, ``stellar mass''
corresponds to the median of the stellar mass probability distribution
marginalised over all other parameters.  At $z<2$, the uncertainties
on the stellar masses are well represented by a Gaussian with
$\sigma=0.04*(1+z)$. However, we note that systematic uncertainties can reach
0.1 dex using different templates or as much as 0.2 dex for massive
galaxies for two different attenuation curves (see Figure 7 in
\cite{2010ApJ...709..644I}).

\subsection{Sample selection}
\begin{figure*}
\begin{center}
  \includegraphics[width=0.49\textwidth]{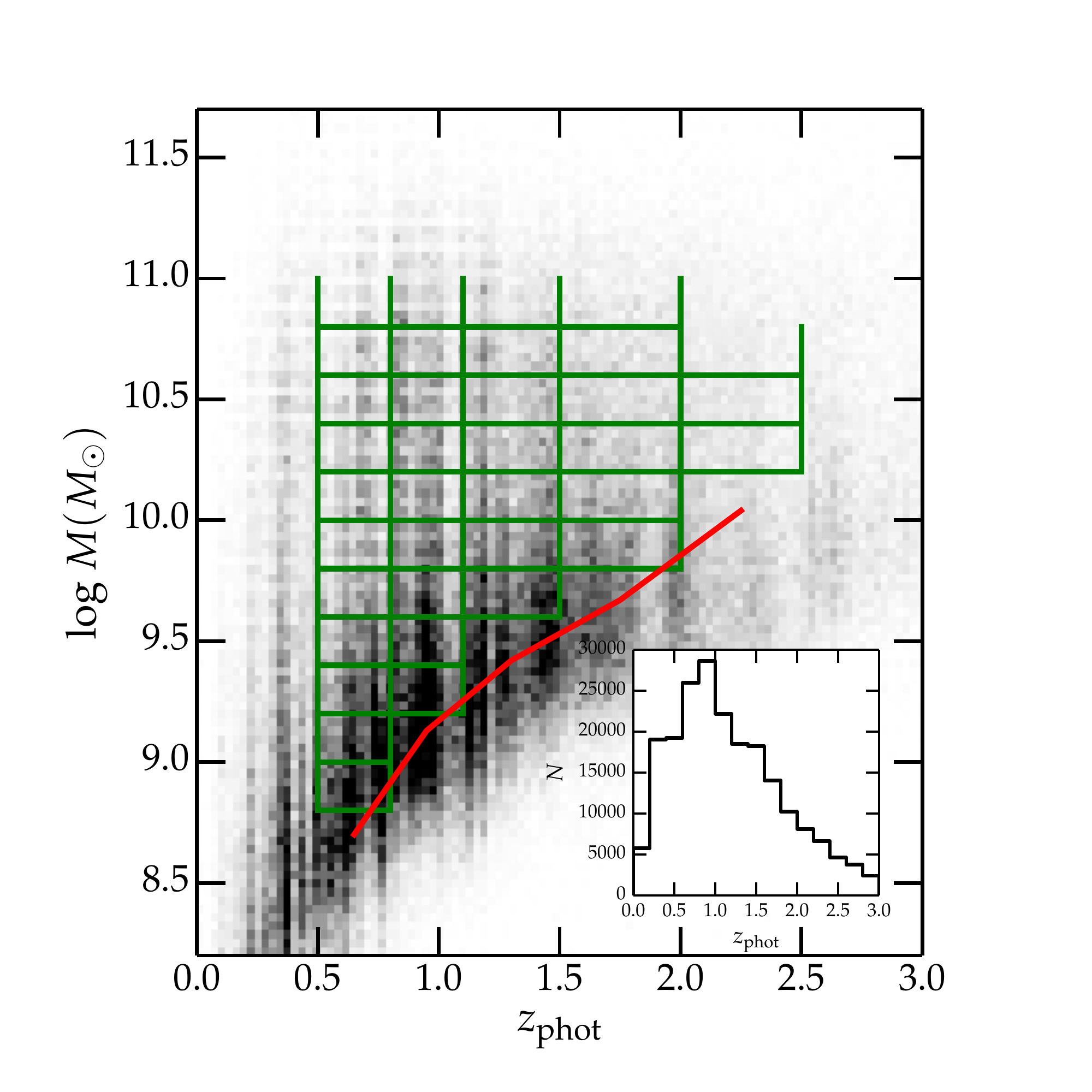}
  \includegraphics[width=0.49\textwidth]{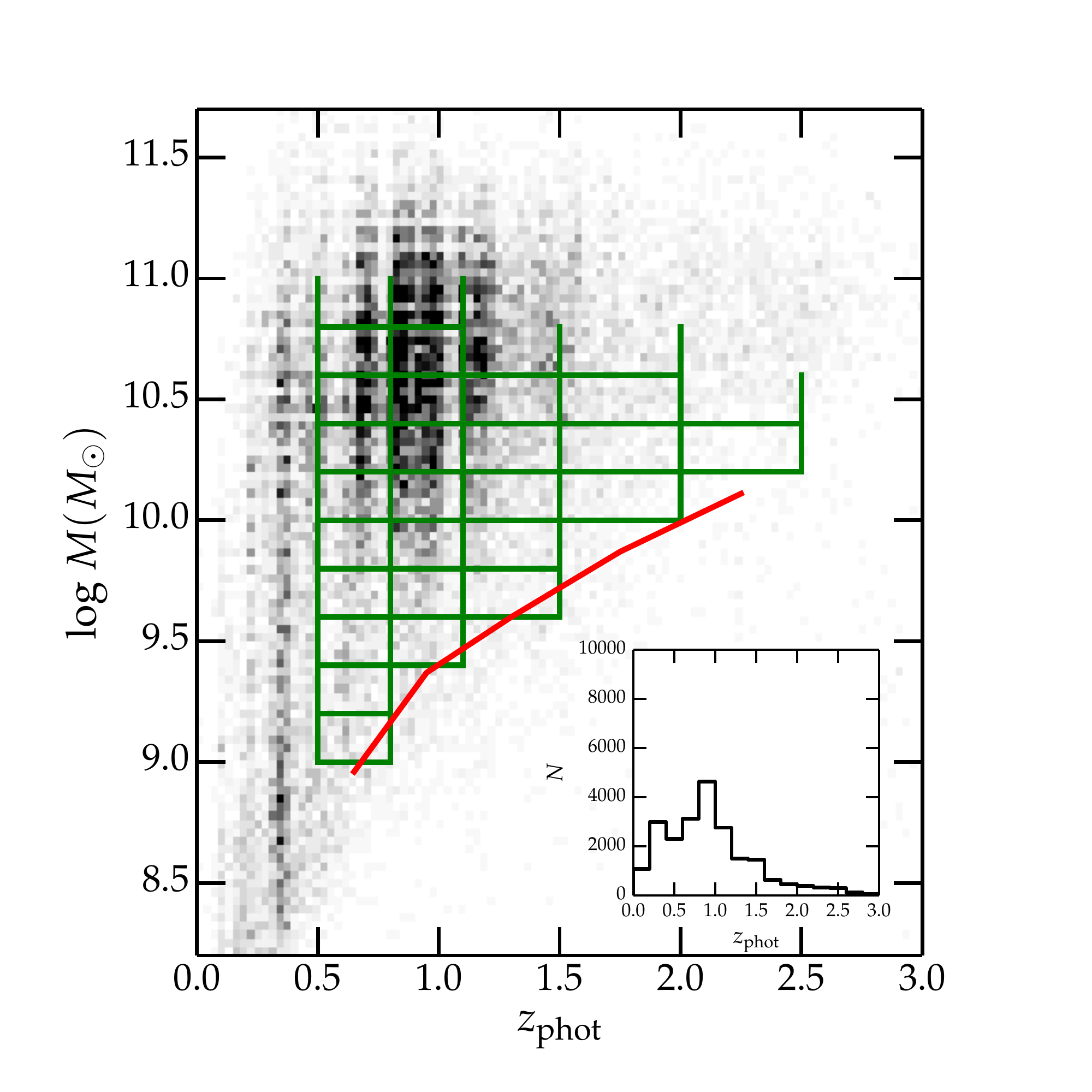}
\end{center}
\caption{The stellar mass-redshift plane for all UltraVISTA galaxies
  (left panel) and for passive galaxies (right panel) with $K_{\rm
    s}<24$. Green lines show our mass thresholds for each sample. The
  solid red line shows the completeness limits from
  \protect\cite{Ilbert:2013dq}. Inset: the redshift
  distributions. (Note: the gray-scale for each pixel in the
  mass-redshift plane is 0--100 objects for the left panel and 0--20 for
  the right panel.) }
\label{fig:selectionVista}
\end{figure*}

We construct a series of volume-limited samples selected by stellar
mass. We first select all galaxies with $K_{\rm s}~<~24$
outside masked regions giving a total of 213,165 objects. After
masking, the field has an effective area of 1.5~deg$^{2}$. The mask
was constructed from a combined COSMOS $B$, $i$ and $V$ mask together
with a mask detailing the borders of the UltraVISTA chi2 image (all
stars are masked when one considers the COSMOS masks).

Figure~\ref{fig:selectionVista} shows the number density of objects as
a function of redshift, with the inset panels showing the redshift
distribution. The red lines are our mass thresholds and the green ones
the completeness limits, as calculated in \cite{Ilbert:2013dq}. To
calculate this completeness limit, \citeauthor{Ilbert:2013dq} computed
the lowest stellar mass which could be detected for a galaxy using the
relation $\log(M_{\rm limit})=\log(M)+0.4\times (K_{s}-24)$ given a
sample at $K_{s}<24$. Then, at a given redshift, the stellar mass
completeness limit corresponds to the mass for which $90\%$ of the
galaxies have their $M_{\rm limit}$ below the stellar mass
completeness limit. The number of objects in each bin, as well as the
mean stellar mass, are summarised in Tables~\ref{tab:smVista} and
\ref{tab:smVistaPassive}. Our large bin widths ($\Delta_{z,min}=0.3$)
ensures a low bin-to-bin contamination but still reduces substantially
the mixing of physical scales at a given angular scale. Our selected
redshift bins are the same as those used in
\citeauthor{Ilbert:2013dq}.

\begin{table*}
\begin{minipage}{126mm}
\begin{tabular}{ccccccccccc}
\hline
&\multicolumn{2}{|c|}{$0.65<z<0.95$}&\multicolumn{2}{|c|}{$0.95<z<1.3$}&\multicolumn{2}{|c|
}{$1.3<z<1.75$}&\multicolumn{2}{|c|}{$1.3<z<1.75$}&\multicolumn{2}{|c|}{$1.75<z<2.25$}\\
Threshold$^{(a)}$&$N_{\rm gal}$ &$M_{\rm med}^{(a)}$&$N_{\rm gal}$ &$M_{\rm med}^{(a)}$ & $N_{\rm gal}$ &$M_{\rm med}^{(a)}$&$N_{\rm gal}$ &$M_{\rm med}^{(a)}$&$N_{\rm gal}$ &$M_{\rm med}^{(a)}$\\
\hline
8.8 & 26441 & 9.43 & -- & -- & -- & -- & -- & -- & -- & -- \\
9.0 & 21642 & 9.60 & -- & -- & -- & -- & -- & -- & -- & -- \\
9.2 & 17300 & 9.81 & 25317 & 9.84 & -- & -- & -- & -- & -- & -- \\
9.4 & 13763 & 10.01 & 20466 & 10.02 & -- & -- & -- & -- & -- & -- \\
9.6 & 10911 & 10.19 & 16431 & 10.21 & 22666 & 10.11 & -- & -- & -- & -- \\
9.8 & 8752 & 10.34 & 13201 & 10.36 & 17361 & 10.29 & 16877 & 10.25 & -- & -- \\
10.0 & 7015 & 10.45 & 10520 & 10.50 & 13280 & 10.45 & 12547 & 10.42 & -- & -- \\
10.2 & 5398 & 10.57 & 8382 & 10.61 & 10000 & 10.58 & 9227 & 10.57 & 5681 & 10.58 \\
10.4 & 3944 & 10.70 & 6258 & 10.74 & 7242 & 10.70 & 6484 & 10.70 & 4087 & 10.71 \\
10.6 & 2556 & 10.85 & 4297 & 10.85 & 4828 & 10.83 & 4247 & 10.84 & 2695 & 10.83 \\
10.8 & 1479 & 11.00 & 2579 & 11.00 & 2698 & 10.98 & 2427 & 10.97 & 1535 & 10.98 \\
11.0 & 742 & 11.15 & 1276 & 11.15 & 1232 & 11.14 & 1094 & 11.13 & -- & -- \\
\hline
\end{tabular}
\label{tab:smVista}
\caption{Characteristics of each redshift bin for the full
  $K_{\rm s}<24.0$ galaxy sample. For each stellar mass threshold and
  redshift bin we report the number of galaxies and the median log
  stellar mass. $^{(a)}$: in $\logm$.}
\end{minipage}
\end{table*}

\begin{table*}
\begin{minipage}{126mm}
\begin{tabular}{ccccccccccc}
\hline
&\multicolumn{2}{|c|}{$0.65<z<0.95$}&\multicolumn{2}{|c|}{$0.95<z<1.3$}&\multicolumn{2}{|c|
}{$1.3<z<1.75$}&\multicolumn{2}{|c|}{$1.3<z<1.75$}&\multicolumn{2}{|c|}{$1.75<z<2.25$}\\
Threshold$^{(a)}$&$N_{\rm gal}$ &$M_{\rm med}^{(a)}$&$N_{\rm gal}$ &$M_{\rm med}^{(a)}$ & $N_{\rm gal}$ &$M_{\rm med}^{(a)}$&$N_{\rm gal}$ &$M_{\rm med}^{(a)}$&$N_{\rm gal}$ &$M_{\rm med}^{(a)}$\\
\hline
9.0 & 3877 & 10.44 & -- & -- & -- & -- & -- & -- & -- & -- \\
9.2 & 3705 & 10.45 & -- & -- & -- & -- & -- & -- & -- & -- \\
9.4 & 3547 & 10.49 & 5526 & 10.59 & -- & -- & -- & -- & -- & -- \\
9.8 & 3186 & 10.55 & 5132 & 10.64 & 3560 & 10.66 & -- & -- & -- & -- \\
10.0 & 2910 & 10.60 & 4775 & 10.67 & 3342 & 10.69 & 1527 & 10.69 & -- & -- \\
10.2 & 2522 & 10.67 & 4299 & 10.73 & 2998 & 10.73 & 1390 & 10.73 & 785 & 10.80 \\
10.4 & 2041 & 10.76 & 3583 & 10.80 & 2585 & 10.79 & 1182 & 10.81 & 714 & 10.83 \\
10.6 & 1440 & 10.88 & 2717 & 10.90 & 2023 & 10.88 & 902 & 10.91 & 576 & 10.91 \\
10.8 & 910 & 11.04 & 1803 & 11.01 & 1266 & 11.01 & 600 & 11.02 & -- & -- \\
11.0 & 498 & 11.19 & 939 & 11.15 & -- & -- & -- & -- & -- & -- \\

\hline
\end{tabular}
\label{tab:smVistaPassive}
\caption{Characteristics of each redshift bin for the passive
   $K_{\rm s}<24.0$ galaxy sample. For each stellar mass
  threshold and redshift bin we report the number of galaxies and the
  median log stellar mass. $^{(a)}$: in $\logm$.}
\end{minipage}
\end{table*}

We also considered the quiescent population selected using the
criterion $M(NUV)-M(R) > 3.5$ defined in \cite{Ilbert:2013dq}. After
applying object masks, the passive sample is composed of 22,169
objects with, as before, a magnitude cut of $K_{\rm s}< 24.0$ . The
characteristics of each sample are summarised in Tables
\ref{tab:smVista} and \ref{tab:smVistaPassive}. The right panels of
Figure~\ref{fig:selectionVista} show the redshift distribution of this
population.

\section{Methods}

\subsection{The angular two-point correlation function}

We measure the two-point angular correlation function $w(\theta)$ for
our samples using the \citet{Landyetal93} estimator, \beq
w(\theta)=\frac{n_{\rm r}(n_{\rm r}-1)}{n_{\rm d}(n_{\rm d}-1)}
\frac{DD}{RR} - \frac{n_{\rm r}-1}{n_{\rm d}} \frac{DR}{RR} + 1,
\label{eq:LS}
\eeq where, for a chosen bin from $\theta$ to $\theta$ +
$\delta\theta$, $DD$ is the number of galaxy pairs of the catalog in
the bin, $RR$ the number of pairs of a random sample in the same bin,
and $DR$ the number of pairs in the bin between the catalog and the
random sample. $n_{d}$ and $n_{r}$ are the number of galaxies and
random objects respectively.  A random catalog is generated for each
sample with the same geometry as the data catalog using
$n_{\rm r} \sim 400,000$ which is at the most more than 500 and at
least 16 times the number of data at each bin.  We measure $w$ in each
field using a fast two-dimensional tree code in the angular range
$0.001 < \theta < 0.2$ degrees divided into 15 logarithmically spaced
bins.

The errors on the two-point correlation measurements are estimated
from the data using the jackknife approach \citep[see for
example][]{Norbergetal11} using 128 subsamples. Removing one
sub-sample at a time, this allows us to compute the covariance
matrix as:
\begin{equation}
C(w_{i},w_{j})=\frac{N-1}{N}\sum_{l=1}^{N} (w_{i}^{l}-\overline{w}_{i})(w_{j}^{l}-\overline{w}_{j}),
\end{equation}
where $N$ is the total number of subsamples, $\overline{w}$ the mean
correlation function and $w^{l}$ the estimate of $w(\theta)$ with the
$l$-th subsample removed.

Finally, although one of the largest survey at these redshifts, the
UltraVISTA field covers a relatively small area, and as a consequence
the integral constraint \citep[see][]{GP77} is expected to have a
impact on our clustering measurements leading on an underestimation
of the clustering strength by a constant factor $w_{c}$ which can be
estimated as follows:
\begin{equation}
w(\theta)=w_{mes}(\theta) + w_{c} .
\end{equation}
Assuming that the two-point correlation function is described by a
simple power with slope $\gamma$ and amplitude $A$ fitted on the data, it
leads to:
\begin{equation}
w_{mes}(\theta)=A\theta^{1-\gamma}-w_{c} \times A(\theta^{1-\gamma}-C).
\end{equation}
We can derive $C$ following \cite{Rocheetal99}:
\begin{equation}
C=\frac{\sum \theta^{1-\gamma}RR(\theta)}{\sum RR(\theta)},
\end{equation}
and then:
\begin{equation}
w(\theta)=w_{mes}(\theta)\frac{\theta^{1-\gamma}}{\theta^{1-\gamma}-C}.
\end{equation}

We find $C \sim 1.42$.

\subsection{Halo model implementation and fitting}
\label{sec:halo-model-impl}

To connect galaxies to their hosting dark matter haloes, we use a
phenomenological ``halo'' model \citep[for a review
see][]{Coorayetal02}. In this model, it is assumed that the number of
galaxies in a given dark matter halo is a simple monotonic function of
the halo mass. By combining this function (the ``halo occupation
distribution'') with our knowledge of the halo mass function and mass
profile, one may predict the abundance and clustering properties of
the visible population.

The key underlying assumption is that the number of galaxies $N$
within a halo depends only on the halo mass $M$ and not on environment
or formation history: we will address the extent to which these
assumptions are reasonable in subsequent sections. 

Our model follows closely \cite{Zhengetal05} who, motivated by
simulations, suggested that the total numbers of galaxies in dark
matter halo, $N(M)$ is a sum of two contributions: one from the
central galaxy in the halo $N_{c}(M)$ and one coming from the
satellites $N_{s}(M)$. Thus $N(M)$ can be expressed as:

\begin{equation}
N(M)=N_{c}(M) \times [1+N_{s}(M)].
\end{equation}
We follow \cite{Zhengetal07} in which the central galaxy is described
as a step function with a smooth transition allowing some scatter in
the stellar mass halo mass relation:
\begin{equation}
N_{c}(M)=\frac{1}{2}\Big[1+\textrm{erf}\Big(\frac{\log M-\log
  M_{min}}{\sigma_{\log M}}\Big)\Big],
\end{equation}
and a power law with a cut at low halo mass for the satellites:
\begin{equation}
N_{s}(M)=\Big(\frac{M-M_{0}}{M_{1}}\Big)^{\alpha}.
\end{equation}
Our model has five adjustable parameters: $M_{\rm min}$, $M_{1}$,
$M_{0}$, $\alpha$ and $\sigma_{\log M}$. In this work,  will examine
in particular $M_{\rm min}$, which represents the characteristic
mass scale for which $50\%$ of haloes host a galaxy, and $M_{1}$,
which is the characteristic mass scale for haloes to host one
satellite galaxy.

Thus the mean number density of galaxies is given by:
\begin{equation}
n_{\rm {gal}}(z)=\int N(M)n(M,z)dM, 
\label{eq:ngal}
\end{equation}
where $n(M,z)$ is the halo mass function for which we use the
prescription from \cite{Sethetal99}.

We use a Navarro-Frenk-White halo density profile
\citep{Navarroetal97} and the halo bias parametrisation $b_{h}(M,z)$
from \cite{Tinkeretal05} which has been calibrated on simulations. 

We compute the following derived parameters:  the mean halo mass:
\begin{equation}
\langle M_{\rm {halo}} \rangle (z)=\int dM M n(M,z)\frac{N(M)}{n_{\rm gal}(z)},
\label{eq:meanhalomass}
\end{equation}
the mean galaxy bias:
\begin{equation}
b_{\rm{gal}}(z)= \int dM b_:{h}(M,z)n(M,z) \frac{N(M)}{n_{\rm {gal}}(z)}
\label{eq:galbias}
\end{equation}
and the satellite fraction:
\begin{equation}
f_{s}(z)=1-f_{c}(z)=1-\int dM n(M,z) \frac{N_{c}(M)}{n_{\rm{gal}}(z)}.
\label{eq:frsat}
\end{equation}

The implementation of the halo model we use is described fully in
\cite{Couponetal12}.

We derive the best-fitting halo models corresponding to our
measurements using the ``Population Monte Carlo'' (PMC) technique as
implemented in the
\texttt{CosmoPMC}\footnote{\texttt{http://cosmopmc.info}} package to
sample likelihood space \citep{Wraithetal09, Kilbingeretal10}. For
each galaxy sample, we simultaneously fit both the two-point
correlation function $w$ and the number density of galaxies $n_{gal}$,
by summing both contributions to the total $\chi^2$:
\begin{eqnarray}
\chi^{2}=\sum_{i,j}[w^{\rm{obs}}(\theta_{i})-w^{\rm{model}}(\theta_{i})](C^{-1})_{i,j}[w^{\rm{obs}}(\theta_{j})-w^{\rm{model}}(\theta_{j})]
\nonumber 
\end{eqnarray}
\begin{eqnarray}
+ \frac{[n^{\rm{obs}}_{\rm{gal}}-n^{\rm{model}}_{\rm{gal}}]^{2}}{\sigma_{\rm{gal}}^{2}},
\end{eqnarray}
where $C$ is the data covariance matrix. The error on the galaxy
number density $\sigma_{gal}$ contains both Poisson noise and cosmic
variance.

\subsection{Estimating the effect of photometric redshift errors on
  clustering measurements}
\label{sec:effect-phot-redsh}

To independently quantify the number of catastrophic photometric
redshift outliers, we analyse the spatial cross-correlation of
galaxies between different redshift bins. The mis-identification of
photo-$z$'s create physical clustering between otherwise un-correlated
bins. We use the pairwise analysis introduced in
\cite{2010MNRAS.408.1168B} , which considers two redshift bins at a
time. With $w_{ij}$ denoting the angular correlation function between
redshifts $i$ and $j$, the following combination of cross- and
auto-correlation function vanishes for all angular scales $\theta_t$,
\begin{align}
d_t = w_{ij}(\theta_t) \left( f_{ii} f_{jj} + f_{ij} f_{ji} \right) \nonumber \\
	- w_{ii}(\theta_t) \frac {N_i}{N_j} f_{ij} f_{ji} \nonumber \\
 	- w_{jj}(\theta_t) \frac {N_j}{N_i} f_{ji} f_{ii} .
\label{eq:dt}
\end{align}
The number of observed galaxies per bin $i$ is $N_i$.  The
contamination fraction of galaxies originating from the redshift range
given by bin $i$, but mis-placed into bin $j$ due to catastrophic
failure is denoted with $f_{ij}$.  This pairwise approach neglects the
contamination from other bins $k \ne i, j$.  Therefore, the fraction
of galaxies correctly identified in bin $i$ is $f_{ii} = 1 - f_{ij}$.
The approximation of the pairwise analysis is valid for contamination
fractions up to 10\%.

We restrict our analysis to non-adjacent redshift bins ($|i-j| \le
1$), since both the large-scale structure and the photo-$z$ dispersion
create correlations between galaxies from neighbouring bins that are
easily larger than 10\%. We follow \citeauthor{Couponetal12} and
calculate the covariance of the data vector $d_t$ using a Jackknife
estimate. As in \citeauthor{Couponetal12}, we neglect the mixed terms
in (\ref{eq:dt}), which correlate different correlation functions,
since these terms are sub-dominant \citep{2010MNRAS.408.1168B}. The
expression for the covariance then becomes
 \begin{align}
 C_{ts} & = \left\langle d_t d_s \right\rangle \nonumber \\
        & = \left\langle w_{ij}(\theta_t) w_(\theta_s) \right\rangle \nonumber
 		\left( f_{ii} f_{jj} + f_{ij} f_{ji} \right)^2 \nonumber \\
        & + \left\langle w_{ii}(\theta_t) w_{ii}(\theta_s) \right\rangle
 		\left( \frac {N_i}{N_j} f_{ij} f_{ji} \right)^2 \nonumber \\
        & + \left\langle w_{ii}(\theta_t) w_{ii}(\theta_s) \right\rangle
 		\left( \frac {N_j}{N_i} f_{ji} f_{ii} \right)^2 .
 \label{eq:cov_ts}
 \end{align}
We calculate a $\chi^2$ null test with $\chi^2 = \vec d_t C^{-1}_{st} \vec d_s$, and fit
the two parameters $f_{ij}$ and $f_{ji}$.

\begin{figure*}
\begin{center}
  \includegraphics[width=0.49\textwidth]{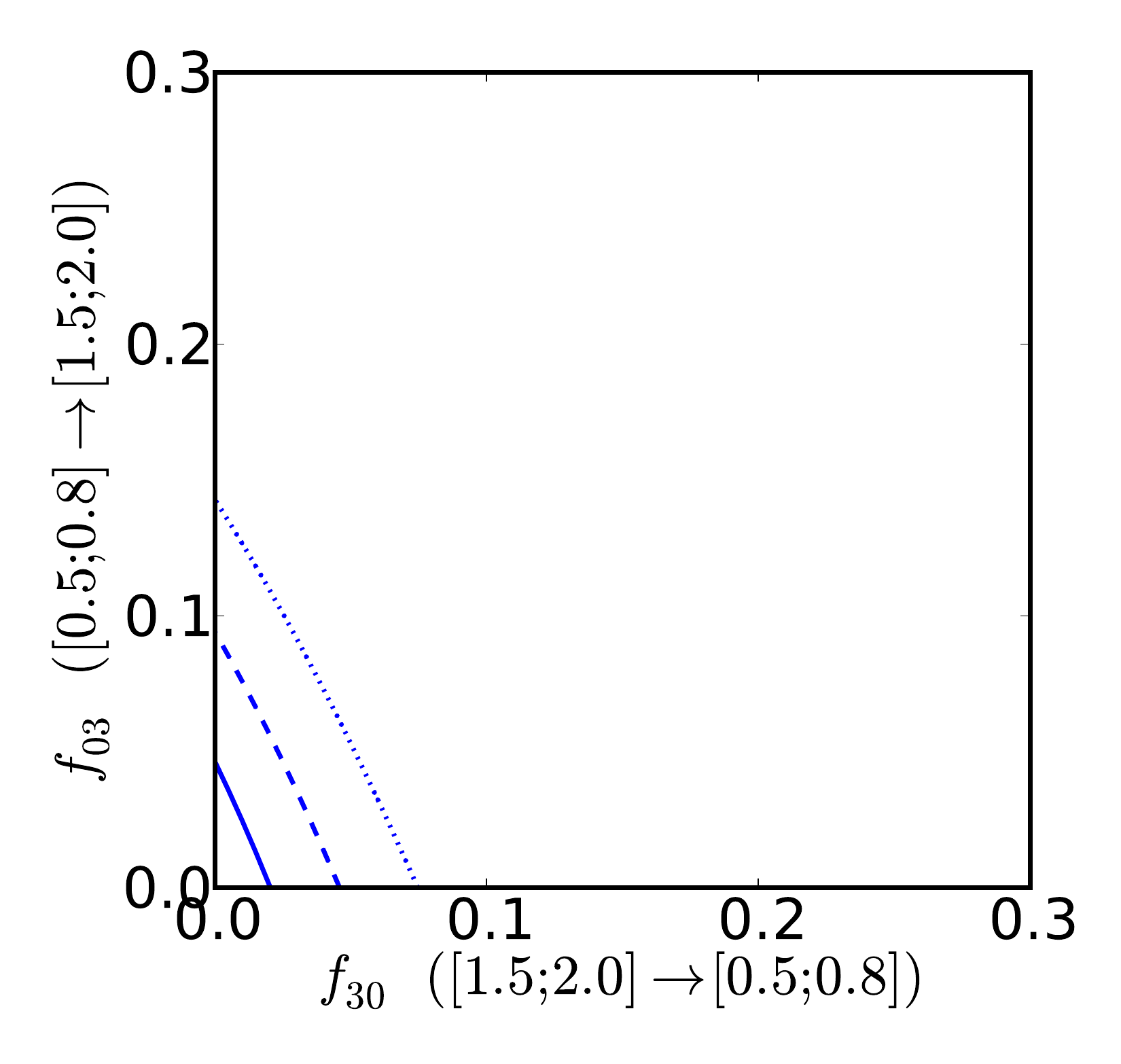}
  \includegraphics[width=0.49\textwidth]{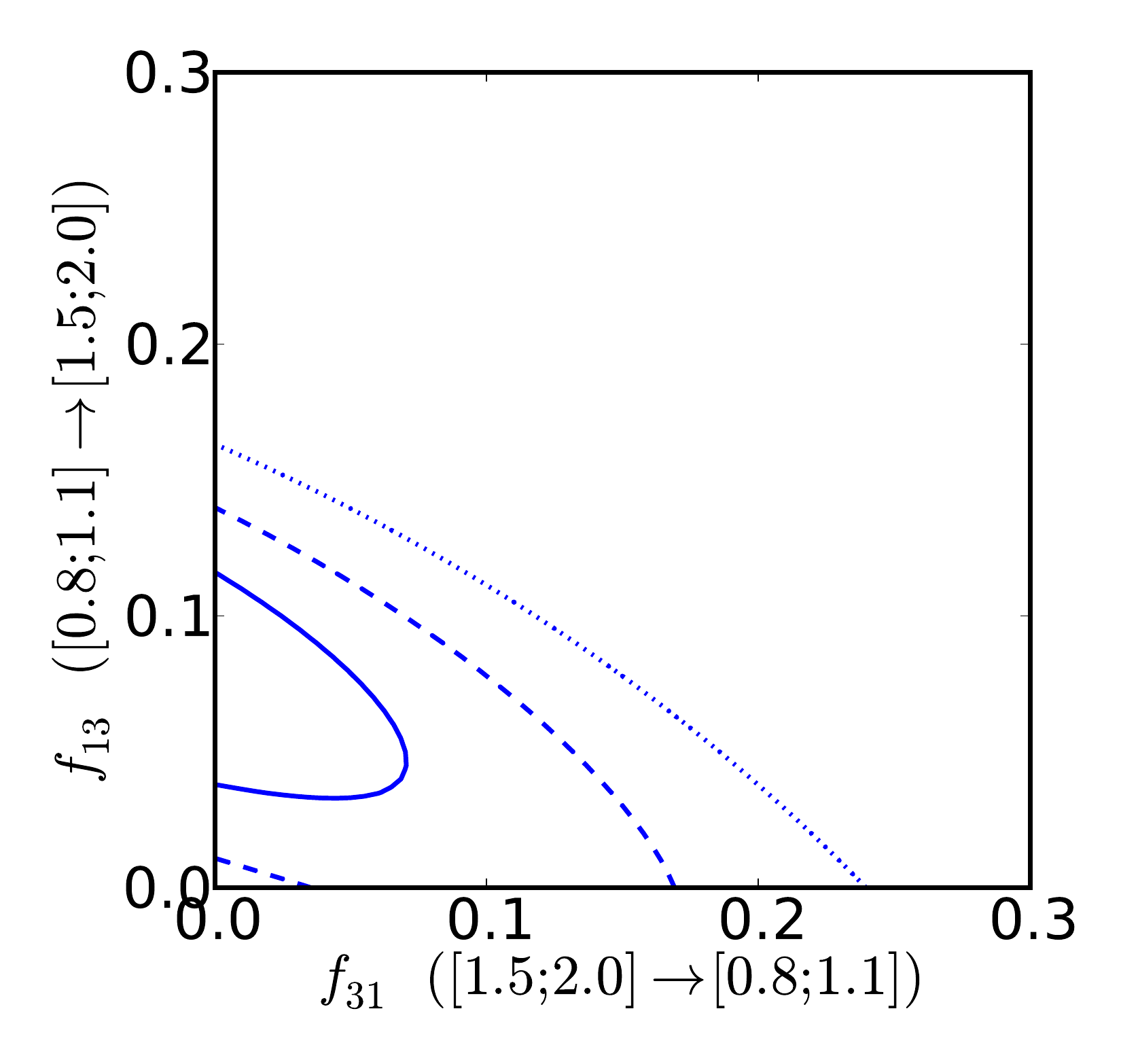}
\end{center}

\caption{Two examples of constraints contamination fraction $f_{ij}$ obtained
from spatial clustering between redshift bin pairs. The solid, dashed, and dotted
lines show $1$-, $2$-, and $3$-$\sigma$ contours, respectively. \emph{Left panel:}
Contamination fraction between bins $[0.5; 0.8]$ and $[1.5;2]$. \emph{Right
panel:} Contamination fraction between bins $[0.8; 1.1]$ and $[1.5;2]$. In both
panels, the $x$-($y$-)axis represents the scattering from high to low (low to
high) redshifts.
}

\label{fig:cross-corr}
\end{figure*}

As an example, we consider the the galaxy sample with
$\logm\sim10.8$. Here, we find the fraction of catastrophic outliers
to be consistent with zero between all pairwise bins (see left panel
of Fig.~\ref{fig:cross-corr} for an example), with the exception of
the contamination $f_{13}$ from bin $0.8 <z < 1.1$ to $1.5 < z< 2.0$,
which is $>3\%$ ($1\sigma$, see right panel of
Fig.~\ref{fig:cross-corr}).

The pairwise analysis typically constrains a quadratic combination of
the contaminations $f_{ij}$ and $f_{ji}$, and does not provide an
independent estimate of the outlier rates.  An upper limit of a
contamination fraction $f_{ij}$ therefore implies that $f_{ji}$ is
zero, or very small. All our upper limits are below $15\%$
($1\sigma$), with the exception of $f_{41}$ and $f_{42}$, for which
the upper limits are $24\%$.  The lowest redshift bin $0.5<z<0.8$ is
affected the least, with contamination fractions less than $8\%$ from
other bins.

To summarise, the cross-correlation analysis independently confirms
the very high quality our of photometric redshifts, and is consistent
with the low catastrophic outlier rate discussed in
Section~\ref{sec:surv-overv-phot}.

\section{Mass-selected clustering measurements}
\label{sec:mass-select-clust}

\begin{figure*}
\begin{center}
  \includegraphics[width=0.49\textwidth]{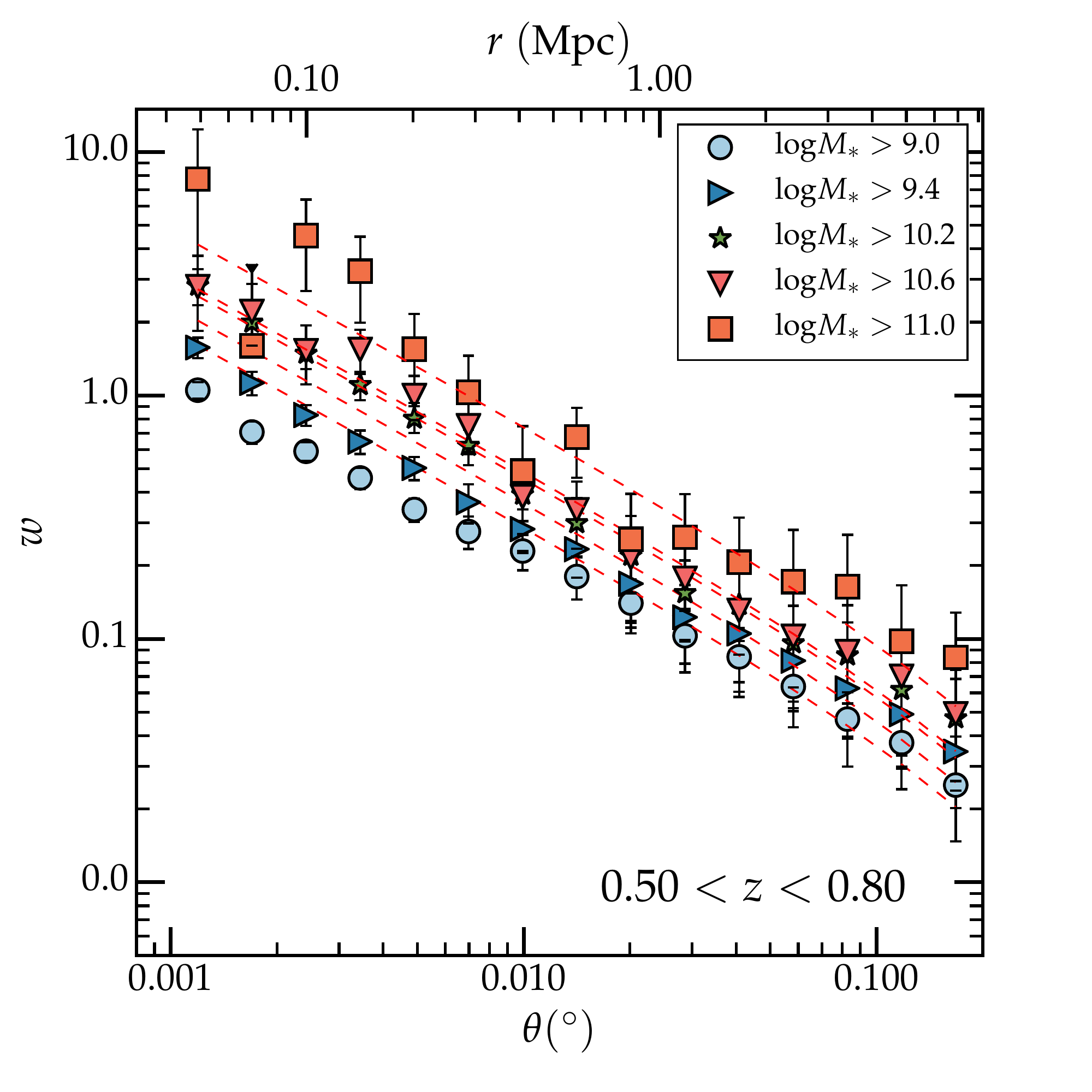}
  \includegraphics[width=0.49\textwidth]{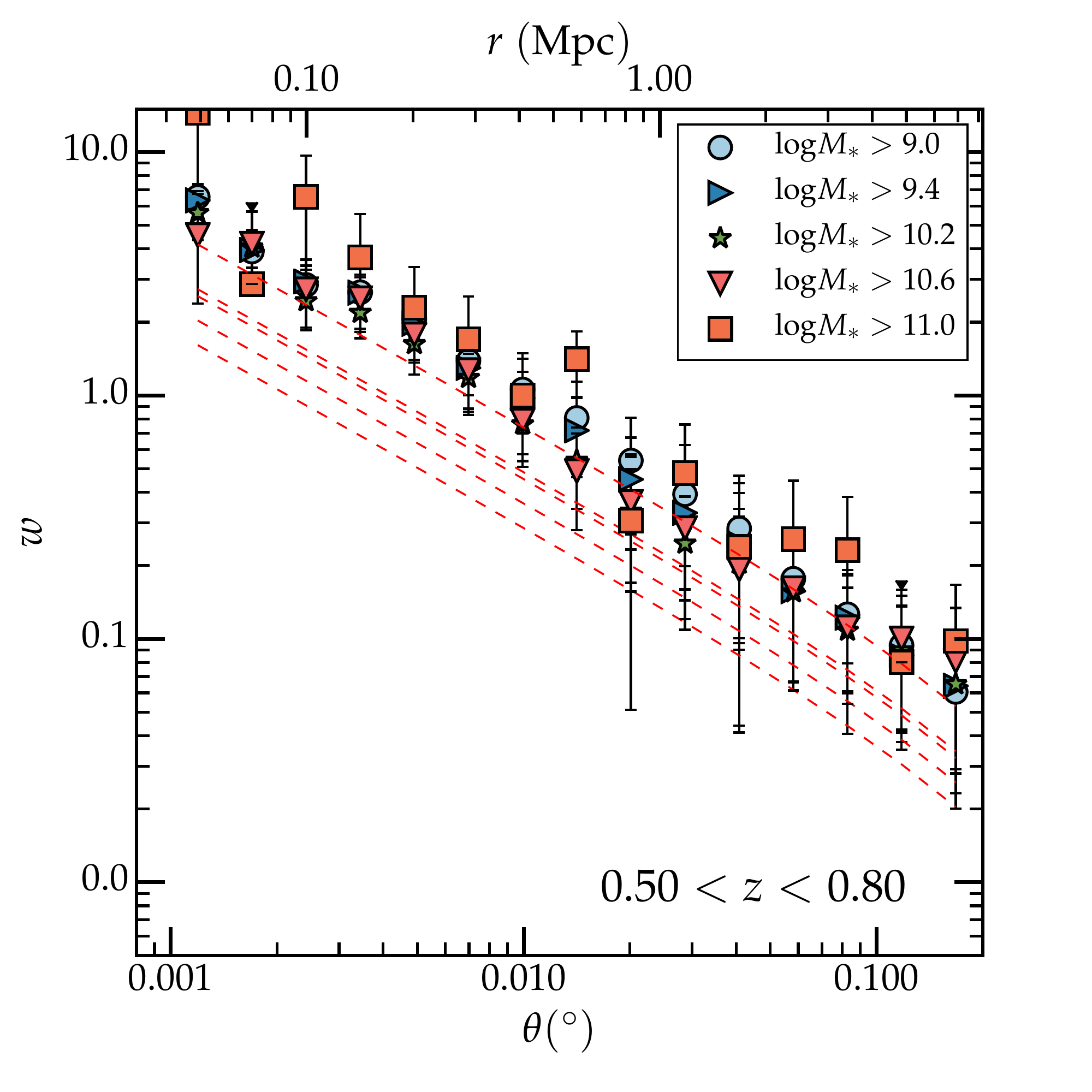}
\end{center}
\caption{Mass-selected galaxy clustering measurements as a function of
  angular scale in degrees in the UltraVISTA-COSMOS survey for the
  full sample (left panel) and the passive galaxy sample (right panel)
  at $0.5<z<0.8$. The dashed lines correspond to the best-fitting
  large-scale power laws for the $0.5<z<0.8$ sample (left panel).}
\label{fig:nofit-1}
\end{figure*}

\begin{figure*}
\begin{center}
  \includegraphics[width=0.49\textwidth]{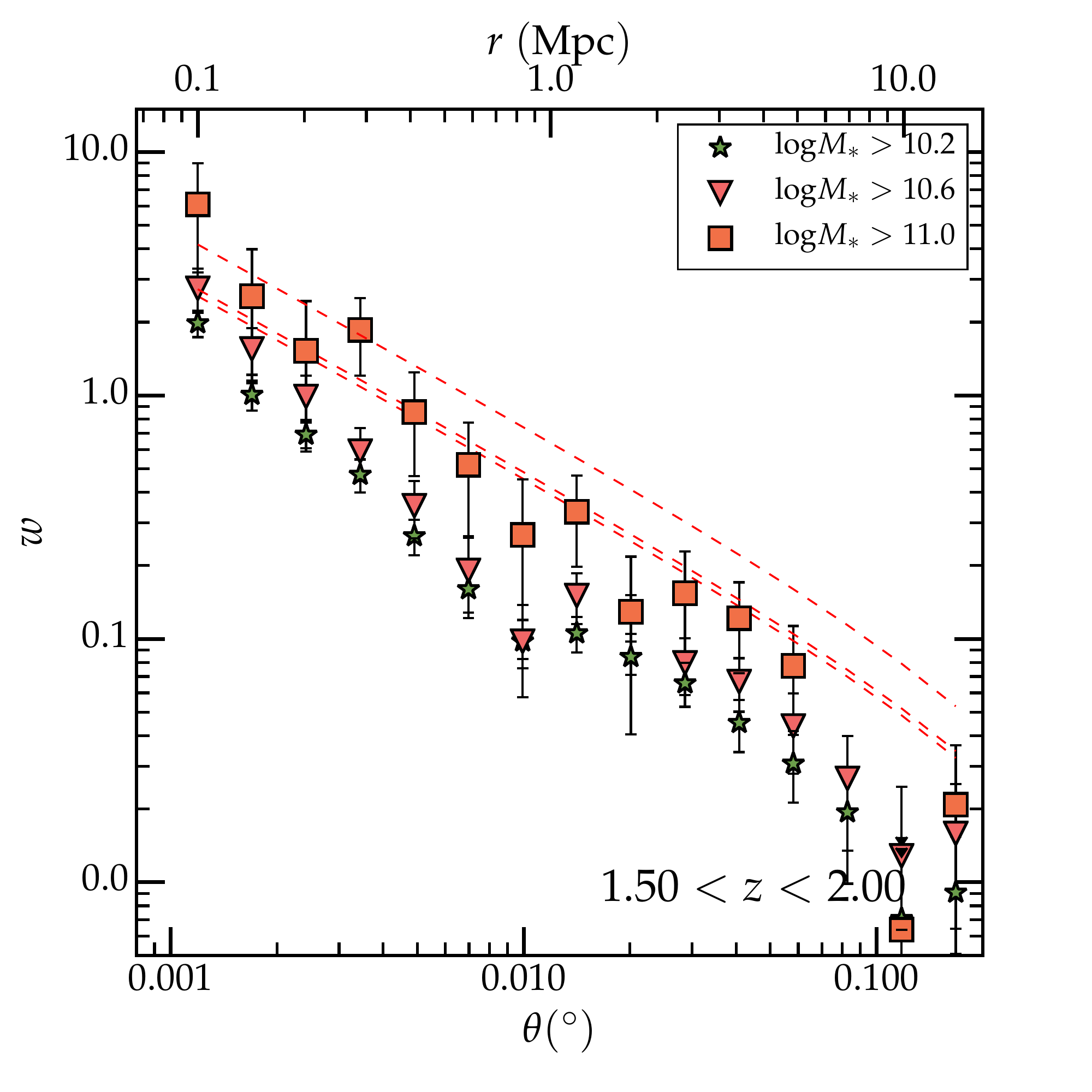}
  \includegraphics[width=0.49\textwidth]{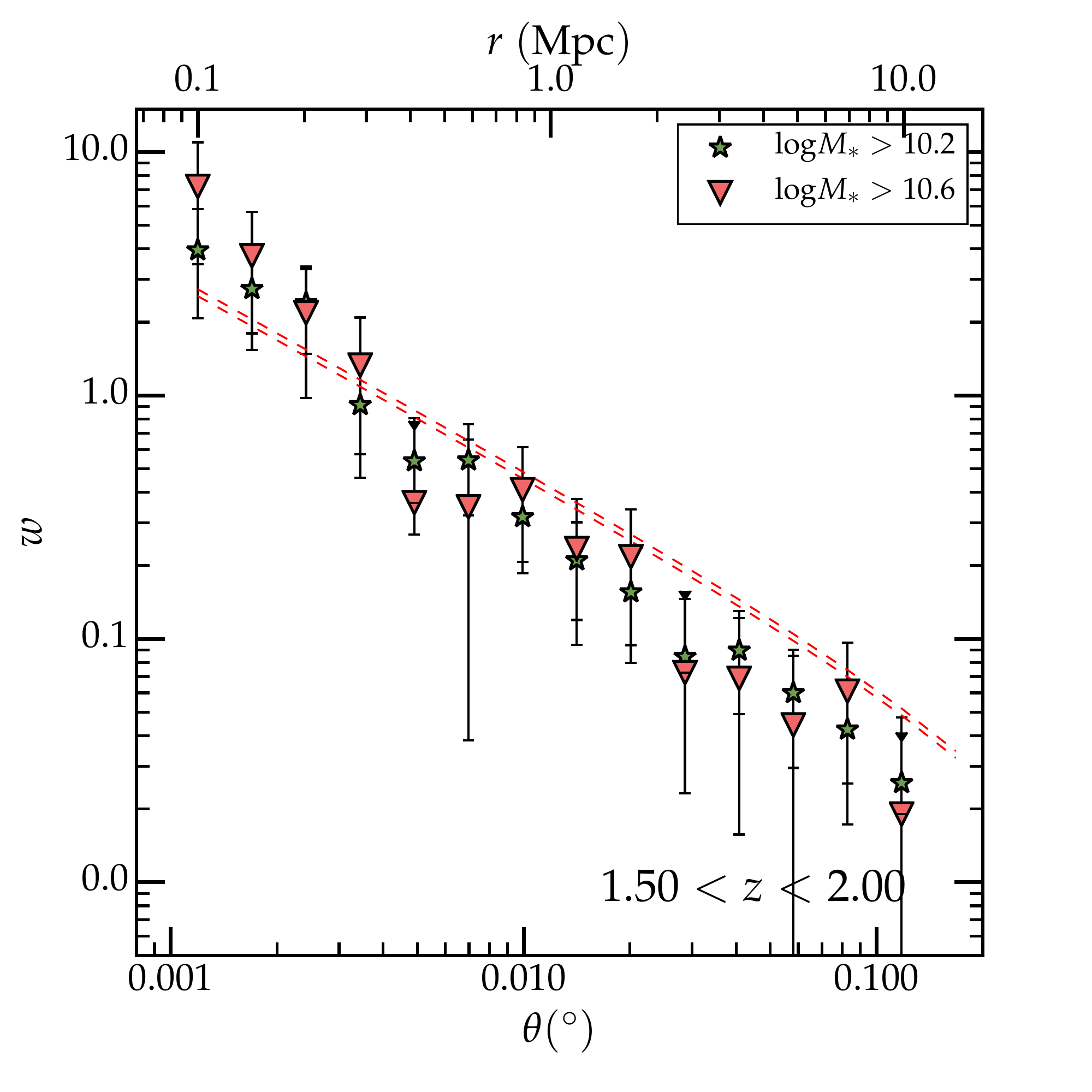}
\end{center}
\caption{As in Figure~\ref{fig:nofit-1} but for the $1.5<z<2.0$
  bin. The dashed lines show the large-scale fit for the corresponding
mass bins in the $0.5<z<0.8$ sample.}
\label{fig:nofit-2}
\end{figure*}

We first consider our mass-selected galaxy clustering measurements. In
Figures~\ref{fig:nofit-1} and \ref{fig:nofit-2} we show the projected
angular correlation function $w$ as a function of angular scale
$\theta$ (in degrees) and stellar mass threshold for two
representative redshift bins, $0.5<z<0.8$ and $1.5<z<2.0$. The left
panels show the full sample, whereas the right panels show the passive
galaxy sample. The dotted lines on all panels correspond to the fits
on large scales ($\sim0.1\deg$) to the low-redshift $(0.5<z<0.8)$ full
galaxy sample with a fixed slope of $-0.8$. Finally, the top
horizontal axes shows the comoving angular separation at the redshift
of the sample.

Qualitatively, several trends are immediately apparent. Firstly, at a
given stellar mass threshold, for both galaxy types, the clustering
amplitude decreases with increasing redshift. Secondly, at a given
redshift, the clustering amplitude is higher for samples with higher
stellar mass thresholds. Finally, at both redshifts and at the
\textit{same} stellar mass threshold, the clustering amplitude of the
passive galaxy population is always higher than the full galaxy
population. It is also interesting to note that the dependence of
clustering strength on stellar mass threshold is less pronounced for
the passive galaxy population at lower redshifts (although this is not
the case in other redshift bins not shown here). The explanation of
this behaviour is quite straightforward: examining
Figure~\ref{fig:selectionVista}, we see that the bulk of the passive
population at $z\sim0.5$ has stellar masses of $10^{10.5}M_{\sun}$: fainter
thresholds do not appreciably change the bulk median stellar mass
threshold and therefore the overall clustering amplitudes rest
unchanged.

Some general comments can also be made concerning the \textit{shape}
of $w$. Firstly, for intermediate stellar mass threshold samples
($M\sim 10^{10}$ $M_{\sun}$) in the lower-redshift $0.5<z<0.8$ bin,
$w$ follows closely a power-law with a slope $\gamma\sim1.8$. However,
at higher stellar mass thresholds, the slope of $w$ begins to steepen,
whereas at lower stellar mass threshold $M\sim 10^{9}$ $M_{\sun}$ the
slope of $w$ is shallower. At high redshifts, finally, the shape of
$w$ deviates from a simple power-law: this is seen most clearly if one
considers the $M\sim 10^{9.8}$ $M_{\sun}$ at low and high redshifts
(filled pentagons in both cases). At high redshifts, a 'break' is
clearly seen at angular scales of $\sim0.01$ degrees, whereas no such
break is visible a lower redshifts.

It is interesting to consider these measurements in the context of
previous clustering and mass-selected clustering measurements in the
COSMOS field. In an early paper, \cite{Meneuxetal09} used the zCOSMOS
10k spectroscopic sample to create a series of mass-selected galaxy
samples covering $0.2<z<0.5$. Despite the use of spectroscopic
redshifts, the comparatively small number of galaxies and the
consequently limited dynamic range (only one decade in stellar mass)
meant that they were not able to detect clearly the trends outlined
here.

Given the complicated nature of the behaviour of $w$ it is clear that
fitting a simple power-law (with a corresponding integral constraint
correction) misses most of these complex features. In the following
Section we will fit our ``halo model'' to these observed correlation
functions and discuss in detail the behaviour of the corresponding
derived parameters as a function of both redshift and stellar mass
threshold.

\section{Halo model analysis}
\subsection{Fitting the two-point correlation function}

\begin{figure*}
\begin{center}
  \includegraphics[width=0.49\textwidth]{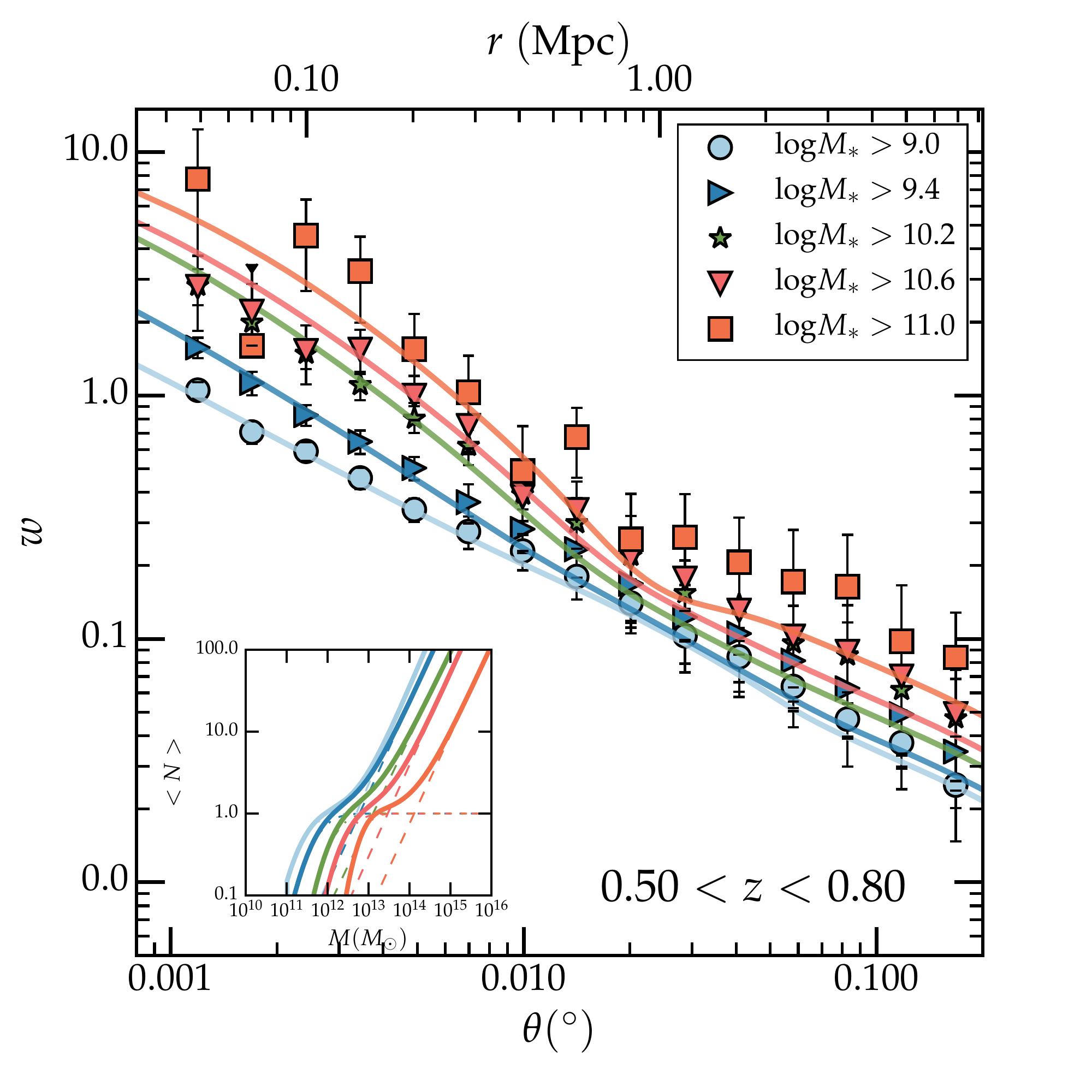}
  \includegraphics[width=0.49\textwidth]{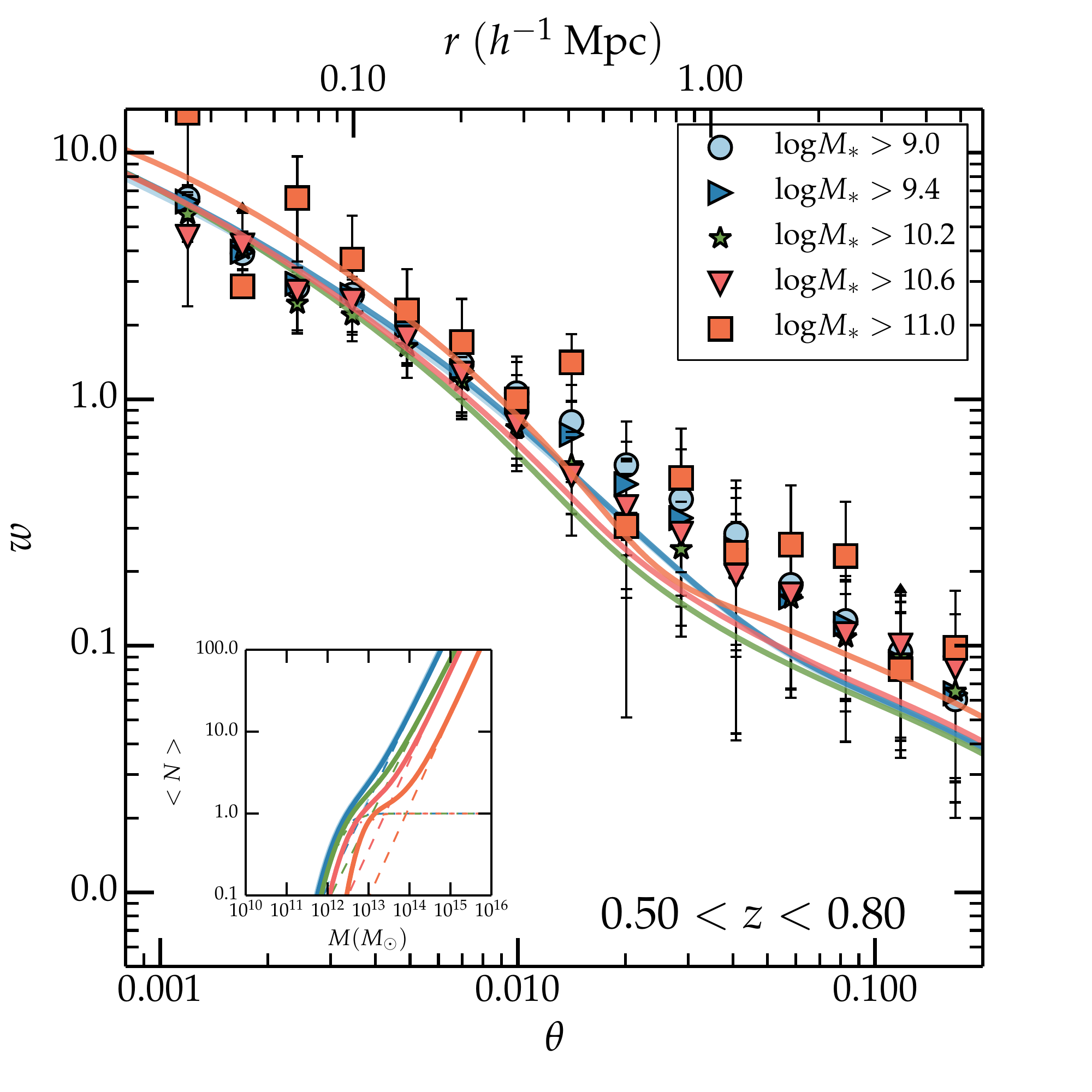}
\end{center}
\caption{Mass-selected galaxy clustering measurements in the
  UltraVISTA-COSMOS survey for the full sample (left panel) and the
  passive galaxy sample (right panel) at $0.5<z<0.8$. The solid lines
  correspond to the best-fitting halo model for each bin. The inset
  panel shows the corresponding halo occupation distribution for each
  of the best-fitting models. Total, satellite and central
  contributions are shown by the solid, dashed and dot-dashed lines
  respectively. The top horizontal axis shows the comoving separation
  corresponding to the angular distance at the effective redshift of
  the slice.}
\label{fig:wVista1}
\end{figure*}

\begin{figure*}
\begin{center}
  \includegraphics[width=0.49\textwidth]{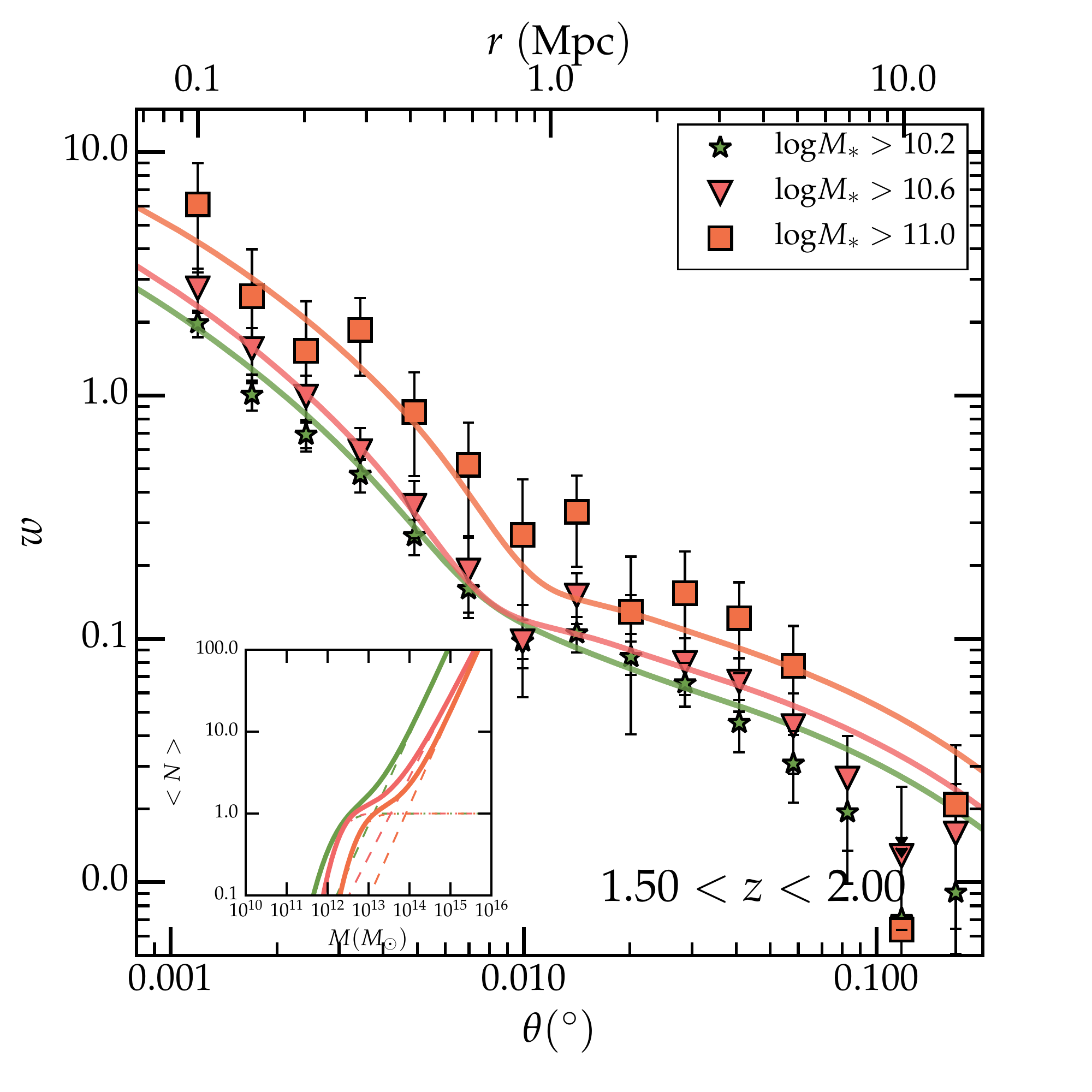}
  \includegraphics[width=0.49\textwidth]{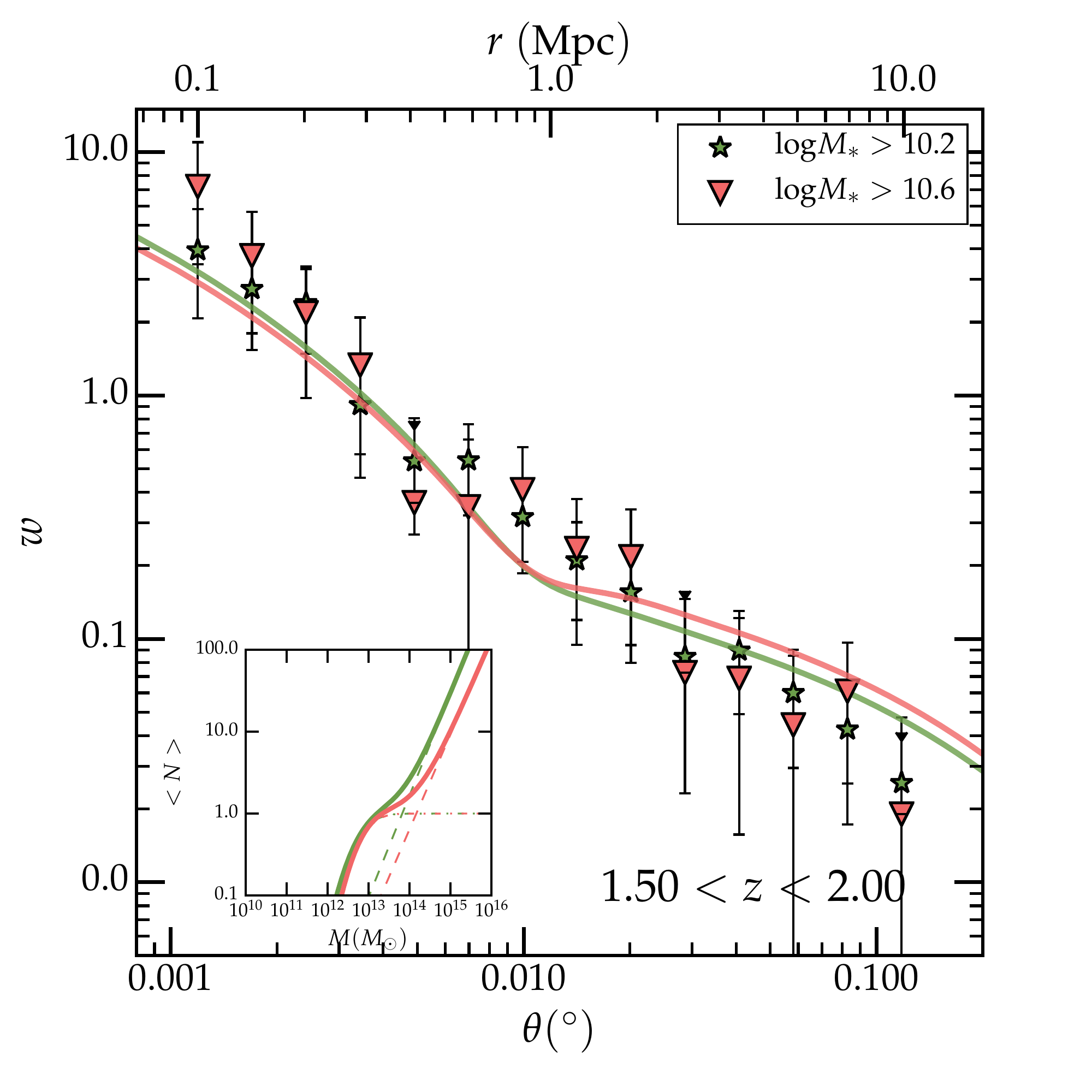}
\end{center}
\caption{As in Figure~\ref{fig:wVista1} but for two samples at $1.5<z<2.0$. }
\label{fig:wVista2}
\end{figure*}

In Figures~\ref{fig:wVista1} and \ref{fig:wVista2} the solid lines
show the best fitting halo model for a range of stellar mass
thresholds for two redshift bins for both passive and total
samples. In each Figure, the thick solid line in the inset panel shows
the corresponding best-fitting halo occupation distribution for each
mass threshold at each redshift. The contribution to the satellite and
central term is shown by the dashed and dotted lines respectively. The
left and right panels show the measurements for the total and passive
sample respectively. 

Qualitatively, the fits are good, in particular for the lower-mass
threshold bins (note, however, the visual inspection of the fits can
be misleading as there is significant co-variance between adjacent
bins). In general, lower
mass threshold bins are better fit by our halo model. In the following
Sections we will consider the derived parameters based on these halo
model fits. 

It is interesting to compare, at the same redshift, the fits for the
passive galaxy population and the full galaxy population. At a
comparable mass threshold, the minimum halo masses are higher (inset
panel on each figure). In addition, it is interesting to note that the
fraction of satellite galaxies is higher for the passive galaxy
sample. We will return to this point in later sections. 

\subsection{Comoving correlation lengths for mass-selected samples}
\label{sec:co-moving-corr}

Traditionally, the comoving correlating length, representing the
amplitude of the real-space correlation function at 1~Mpc and denoted
by $r_0$ has been used as a measure of the strength of galaxy
clustering. Often, this amplitude has been estimated by fitting a
power-law correlation function to the projected correlation function
and using it to estimate (sometimes by extrapolation) the correlation
function amplitude at $1~$Mpc after de-projection using the
\cite{Limber:1954p11344} formula. This procedure can potentially be
problematic: as we have seen, the correlation function is poorly fit
by a simple power law, and often the fitted scales lie outside the
range of the survey. In this work we adopt a different approach: by
using our fitted halo model parameters, we can \texttt{directly}
compute $\xi(r)$, the real-space correlation function, at each slice,
and from this make a direct measurement of the value of the
correlation amplitude at $1~$~Mpc. From the top axes of
Figures~\ref{fig:wVista1} and \ref{fig:wVista2} we note, furthermore,
that this $1~\mathrm{Mpc}$ scale falls within the survey area at all
redshifts.

\begin{figure}
\begin{center}
  \includegraphics[width=0.49\textwidth]{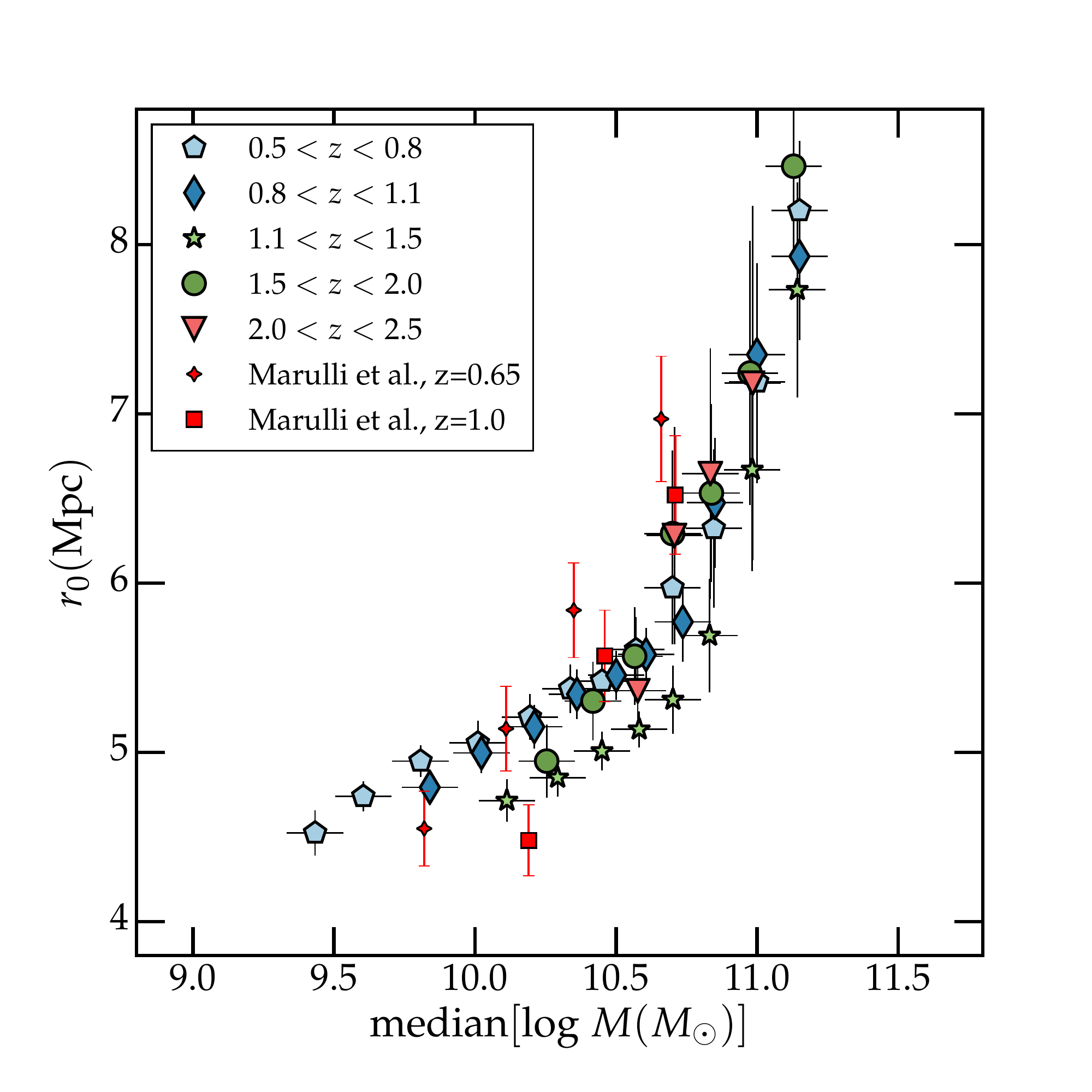}
\end{center}
\caption{The co-moving correlation length $r_0$, in Mpc, computed from
  our halo model for each redshift slice as a function of stellar mass
  threshold. The small squares and diamonds show correlation
  lengths measured in samples thresholded in stellar mass in the
  VIPERS survey (taken from Table 3 in Marulli et al.~2013).}
\label{fig:r0}
\end{figure}

These fits are plotted in Figure~\ref{fig:r0} which shows the comoving
correlation length $r_0$ as a function of sample median stellar
mass. Error bars are computed by measuring the standard deviation of
$r_0$ over a weighted set of 5,000 PMC realisations of our halo model
fits. For reference, small symbols show the values derived by
\cite{Marullietal13} in VIPERS, and within the error bars,
our measurements are in agreement with this work. 

We see that the amplitude of the co-moving correlation length
increases gradually for samples whose mean stellar masses are smaller
than $\logm\sim 11.0$; for samples more massive than this, the
amplitude increases steeply. The presence of this ``knee'' amplitude
has been seen previously in lower redshift samples, at least for
luminosity-selected samples (see, for example
\cite{Norberg:2002p402}). We also we also see that at \textit{fixed}
stellar mass threshold, the clustering amplitude is independent of
redshift. Some hints of this behaviour has been seen in previous
papers \citep{2008A&A...479..321M,Polloetal06,Meneuxetal09}, but this
is the first time it has been unambiguously detected over such a large
redshift range. Finally, we note that the bin $1.1<z<1.5$ is offset
from the others: as we shall see in Section~\ref{sec:repr-cosm-field},
this a consequence of the rich structures present at intermediate
redshifts in the COSMOS field.

\subsection{The characteristic halo mass --  galaxy number density relationship}
\label{sec:ngalM}

\begin{figure}
\begin{center}
  \includegraphics[width=0.49\textwidth]{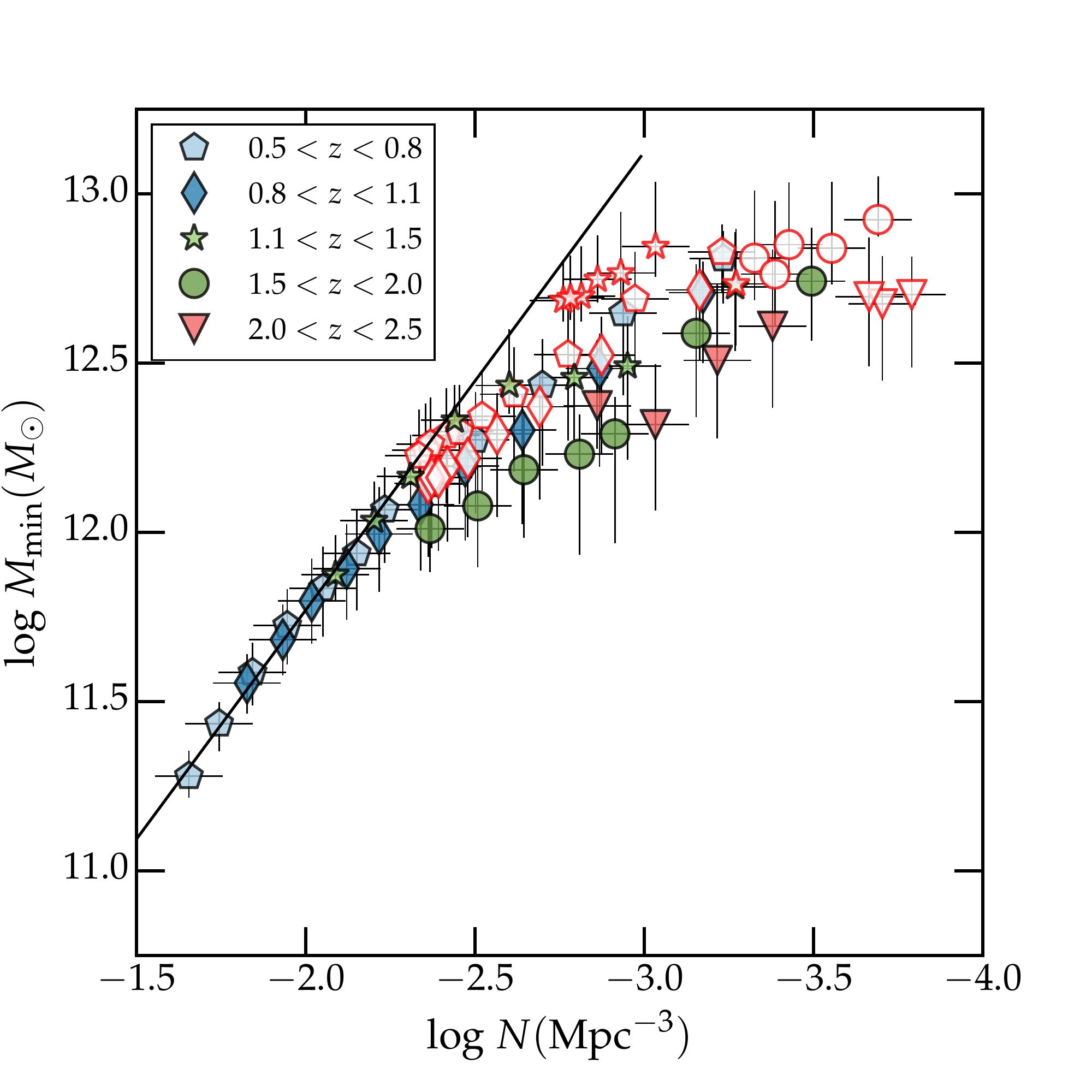}
\end{center}
\caption{$M_{\rm min}$ as a function of log galaxy number density for
  each redshift and mass threshold slice (passive galaxy samples are
  shown by the red open symbols). The solid line corresponds to a fit
  to the low-mass end of the most abundant samples.}
\label{fig:Mhngal}
\end{figure}

Figure~\ref{fig:Mhngal} shows $M_{\rm min}$ as a function of galaxy
number density (defined in Equation~(\ref{eq:ngal}) for the full and
the quiescent samples (open red symbols respectively). In general,
rarer, less abundant objects reside in more massive haloes. Comparing
the quiescent population with the full galaxy sample, we see that
(within the error bars) for a given galaxy abundance both the
quiescent and full galaxy populations lie within haloes of the
\textit{same} dark matter haloes masses (with the exception of the
$1.1<z<1.5$ bin, to which we will return to later; this bin is
systematically different from all the others). In other words, in the
halo mass / stellar mass plane, nothing distinguishes the passive
population from the full galaxy population. 

The solid line shows a power-law fitted on the five most abundant bins
of the lowest-redshift sample $0.5<z<0.8$. It is clear that
\textit{even for a given redshift slice}, a simple power-law fit does
not adequately describe the data. Both the $0.5<z<0.8$ and $0.8<z<1.1$
redshift bins, which have sufficient depth to cover a large range in
abundances show an inflection point at
$\mathrm{log}\sim-2.5$. Higher-redshift bins do not have sufficient
depth to reach below this inflection point, so we cannot say
definitively if this feature is also present in the higher-redshift
data. Concerning redshift evolution of this relation, although our
volume-limited samples cover different mass ranges at different
redshifts, there is some tentative evidence that at fixed abundances,
minimum halo masses required to host galaxies are progressively lower
at higher redshifts (the points at $1.5 <z <2.0$, for example, are
below all the low-redshift points, and this trend continues to even
higher redshifts).

Some previous authors have also considered this
relationship. \citeauthor{Couponetal12}, in the CFHTLS, found no
evidence for an inflection point in the $M_{\rm min}$ versus $n_{\rm gal}$
relationship between $z=0.2$ and $z=1.2$. However, it should be noted
that their samples were only approximately mass-limited; our slope in
Figure~\ref{fig:Mhngal} is steeper than they found. In later Sections
we will discuss how this change in slope is related to the evolution
of the global stellar mass function, and how the origin of the
inflection point is connected to the 

\subsection{Characteristic halo mass scales as a function of stellar mass
  and redshift}
\label{sec:char-halo-mass}

\begin{figure*}
\begin{center}

  \includegraphics[width=0.49\textwidth]{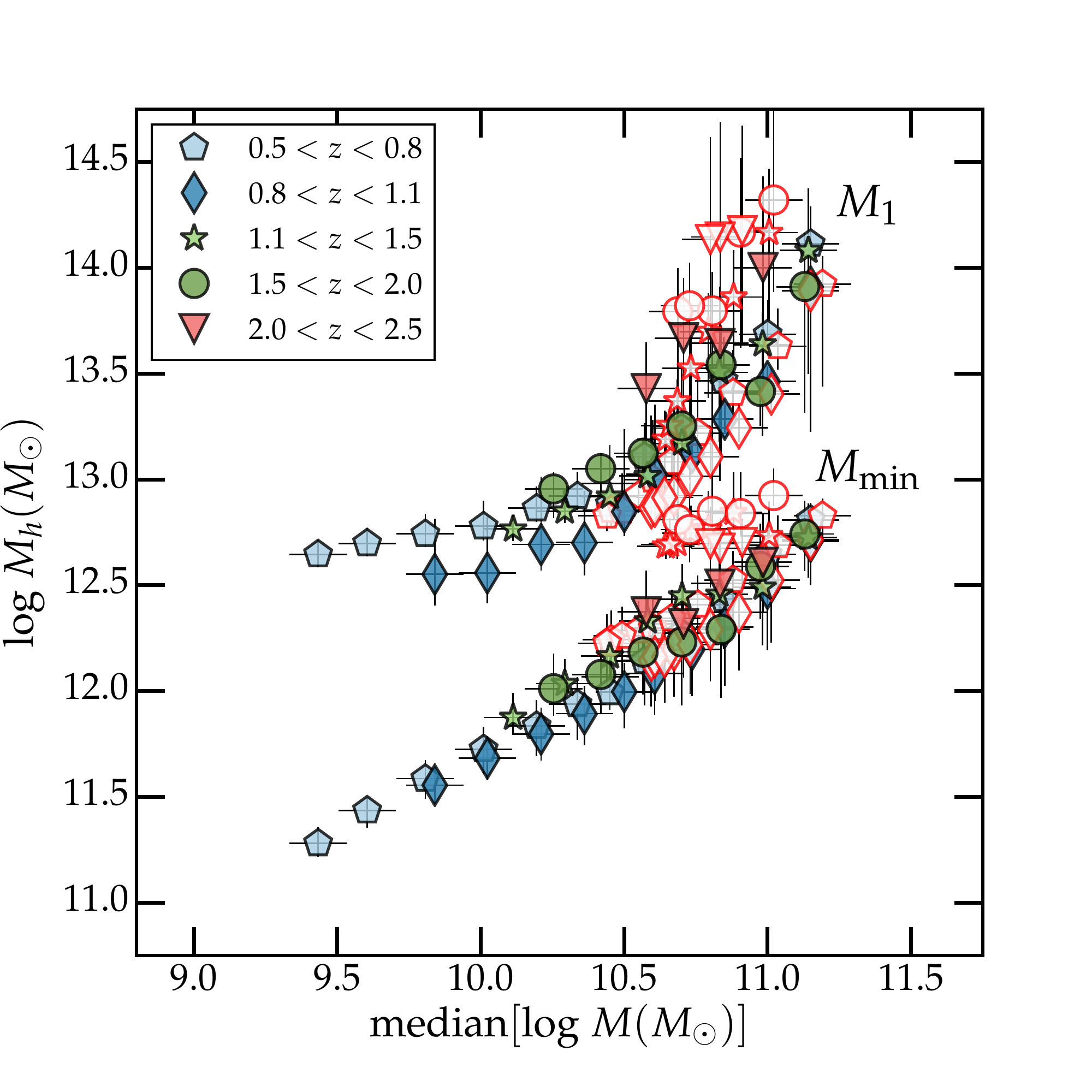}
  \includegraphics[width=0.49\textwidth]{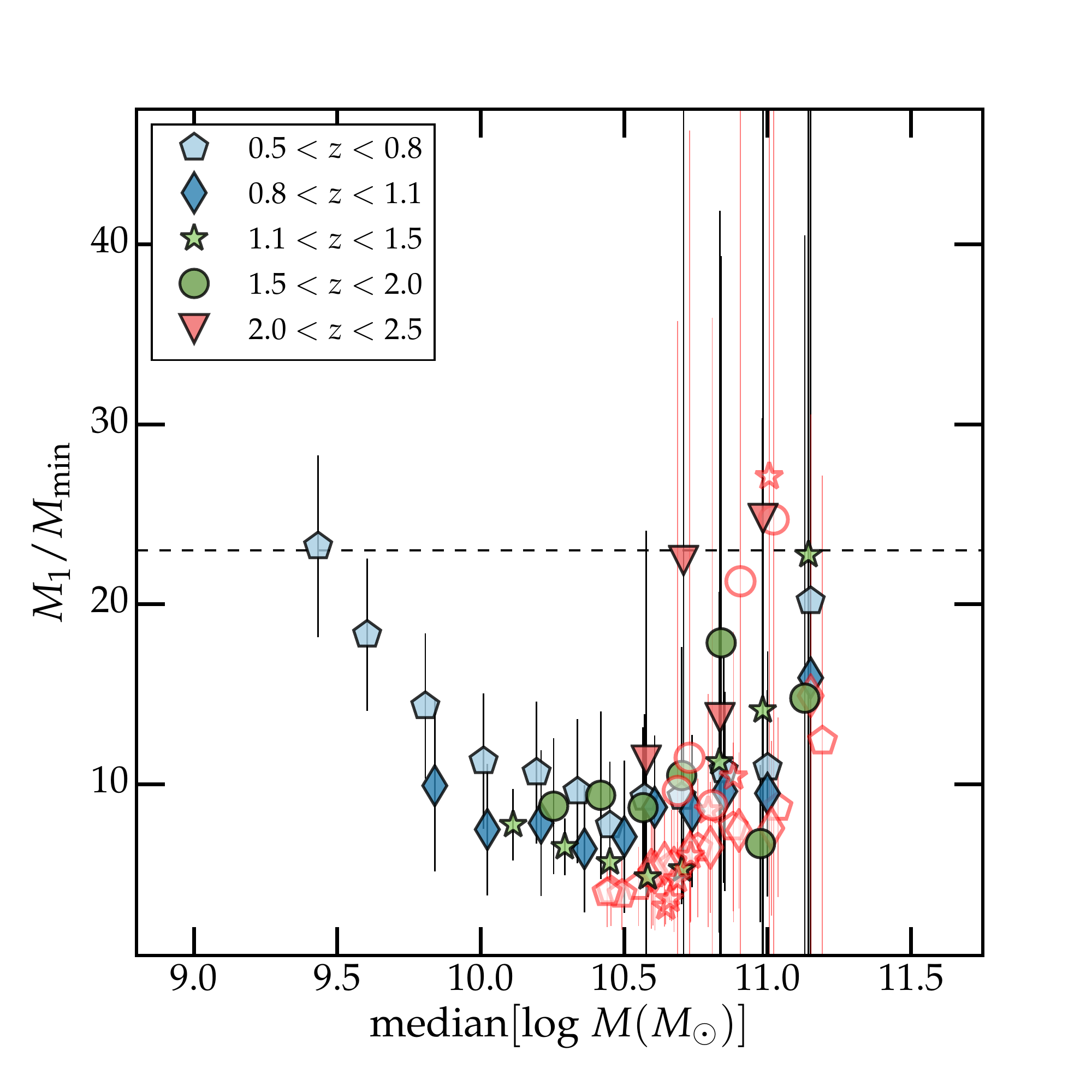}
\end{center}
\caption{Characteristic halo masses $M_{\rm min}$ and $M_{\rm 1}$
  (left panel) and the ratio $M_{\rm min}/M_1$ (right panel) as a function of sample
  median stellar mass threshold. Passive samples are shown in red open
symbols.}
\label{fig:MstarvsMminM1}
\end{figure*}

We now consider the characteristic mass scales $M_{\rm min}$ and
$M_{1}$, representing the minimum halo mass required to host one and
two galaxies respectively. These quantities are shown in the left
panel of Figure~\ref{fig:MstarvsMminM1} as a function of median sample
stellar mass for each redshift bin and mass threshold (as before, red
symbols represent passive galaxy samples). As we have seen in
Figures~\ref{fig:wVista1} and \ref{fig:wVista2}, galaxies with higher
stellar masses reside in progressively more massive dark matter
haloes. In the log-log plane of Figure~\ref{fig:MstarvsMminM1}, this
is an approximately linear relationship with one important exception:
the lowest-mass bin in $M_1$, which flattens out at lower-mass
thresholds. Some hint of this is also seen in the next-nearest mass
threshold, suggesting that this is a generic feature of the lower-mass
threshold samples. There is some evidence in
Figure~\ref{fig:MstarvsMminM1} that, at a fixed stellar mass
threshold, at low redshifts, both $M_{\rm min}$ and $M_{1}$ do not
evolve: however, at $z \sim 1$ they increase sharply with redshift, as
can be seen for the highest redshift bin $2.0<z<2.5$.

We now consider the ``mass gap'' between $M_{\rm 1}$ and $M_{\rm
  min}$: the right panel of Figure~\ref{fig:MstarvsMminM1} shows the
ratio $M_{\rm min}/M_1$. It is useful to first consider the lowest
redshift bin, $0.5<z<0.8$, as this probes the largest stellar mass
thresholds. We can clearly see that this ratio passes through a
minimum at intermediate mass thresholds. For both low-mass and
high-mass stellar masses, this ratio is $\sim 20$; at intermediate
stellar mass thresholds, the ratio is $\sim10$. This allows us to
understand measurements in the literature: at high thresholds in
absolute magnitude (corresponding to our most massive samples),
\cite{Zehavietal11} using SDSS observations at $z \sim 0.1$ found
$\sim 20$; on the other hand, \cite{Wakeetal11} in the NEWFIRM Medium
band survey (NMBS) at $z \sim 1.5$ found much smaller values, $\sim
10$; however as we can see from Figure~\ref{fig:MstarvsMminM1} this is
primarily because these observations probed a much smaller range in
stellar mass thresholds; in Figure~\ref{fig:MstarvsMminM1}, most of
our observations are at this stellar mass threshold.

One interpretation of our results is that at high stellar mass
thresholds, it becomes more difficult (it requires a more massive
halo) to form satellites as the material preferentially falls onto
the central object. There is some evidence also that that ratio
between $M_{1}$ and $M_{\rm min}$ decreases towards higher redshift as
a consequence of the fact $M_{1}$ evolves less rapidly than
$M_{\rm min}$, although our error bars are large in the high redshift
/ high stellar mass bins. \cite{Kravtsovetal04} used high-resolution
dissipationless $N$-body simulations to investigate the halo
occupation distribution and predicted that $M_{1}/M_{\rm min}$ should
have $2/3$ of its $z=0$ value by $z=1$. This prediction is consistent
with what we find between our redshift bins $0.5<z<0.8$ and
$1.1<z<1.5$.  This means that, at higher redshifts, the difference
between haloes containing several galaxies or only one becomes smaller
which could be seen as a evidence that, at higher redshift, haloes may
have more recently accreted satellites.

\subsection{The stellar-mass halo-mass relationship and comparisons
  with abundance-matching measurements}
\label{sec:stellar-mass-halo}

\begin{figure*}
\begin{center}
  \includegraphics[width=\textwidth]{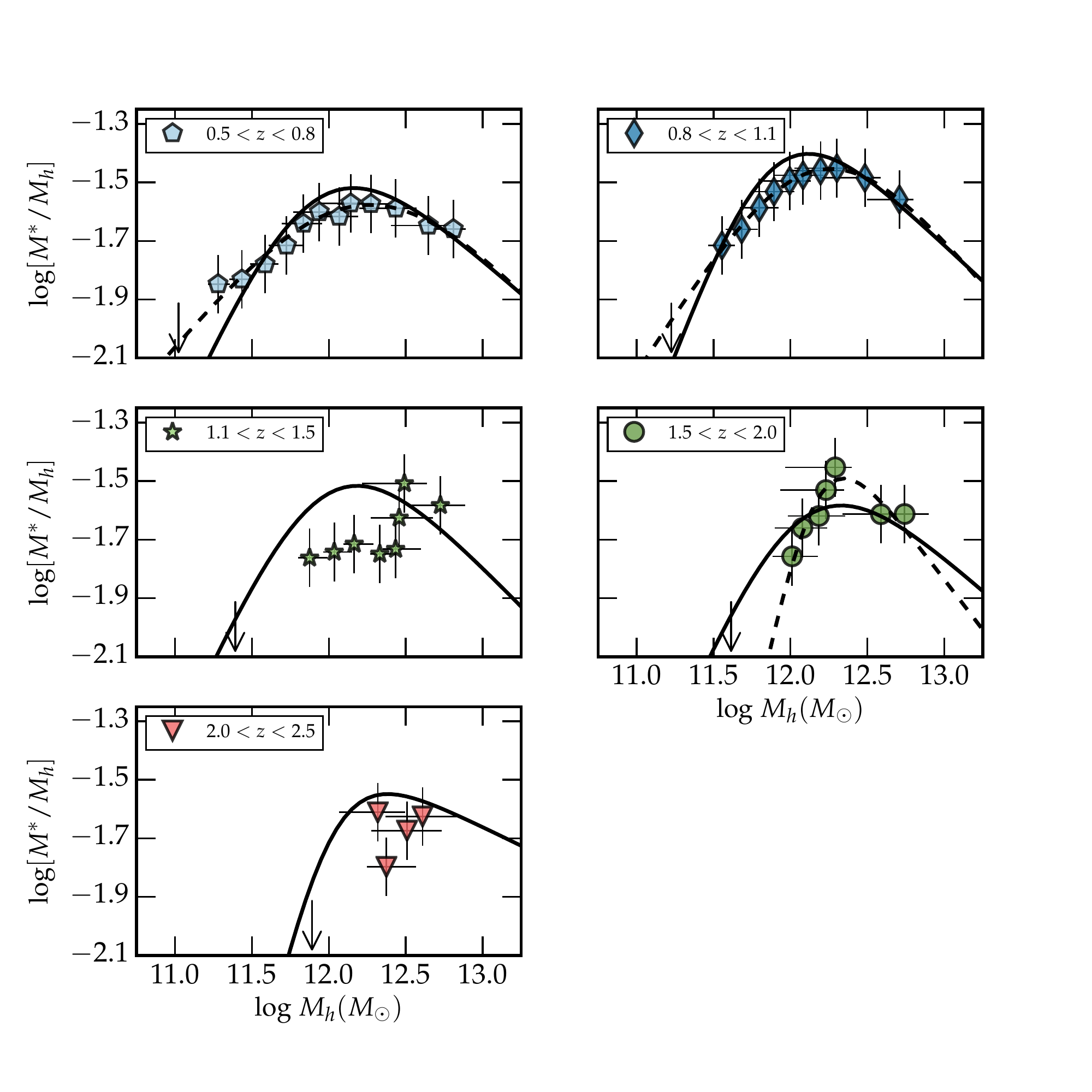}
\end{center}
\caption{The ratio between the median stellar mass in each sample and
  the halo mass at each redshift slice (filled coloured symbols). The
  dotted and solid lines shows fit of the
  \citeauthor{2003MNRAS.339.1057Y} analytic expression to the HOD
  measurements and the abundance matching results. The
  downward-pointing arrow in each redshift slice shows our approximate
  completeness limit in stellar mass, translated to the corresponding
  halo mass at that redshift. }
\label{fig:shmr}
\end{figure*}

Previously, we have considered the relationship between the
characteristic halo mass scales and each samples' stellar mass
threshold. Another way to consider this relationship is to compute the
ratio of stellar mass to halo mass as a function of either halo mass
or stellar mass, known as the stellar-mass halo mass relationship, or
SHMR. This has the advantage of explicitly showing what fraction of
mass in stars is contained within a dark matter halo of given halo
mass. One may then attempt to interpret this quantity in terms of the
integrated star-formation history over the lifetime of the halo, and
in particular the star-formation rate per stellar mass, or the
specific star-formation rate.  The implication is that present-day
haloes which have a higher stellar mass to halo mass relationship are
those in which star-formation was more efficient than the past.
Figure~\ref{fig:shmr} shows, for each redshift slice, the ratio of the
median stellar mass to the characteristic halo mass $M_{\rm {min}}$ as
a function of halo mass.

We fit this ratio to the widely-used relationship of
\cite{2003MNRAS.339.1057Y} which models the SHMR as a double power-law
with a different slope at high-mass and low-mass sides. Although it is
has been suggested that this functional form may not be an optimal
description of the SHMR \citep{Leauthaud:2011bj} we consider it
sufficient for this current dataset, given the uncertainties which
exist concerning the nature of dark matter haloes at high and
intermediate redshifts which are currently not well constrained. The
dashed lines show the fit to the \citeauthor{2003MNRAS.339.1057Y}
relationship, where each point was weighted by the corresponding error
in $M_{\rm min}$ computed by PMC fitting procedure.

For most redshift bins, our COSMOS-UltraVISTA survey provides enough
low mass and high-mass haloes to constrain the SHMR on both sides of
the peak. However, for the $2<z<2.5$ bin we are not able to determine
the peak location, given the challenging nature of correlation
function measurements over sufficiently large stellar mass range at
these redshifts. This is also the case for our ``outlier'' bin
$1<z<2$ (which we already mentioned in Section~\ref{sec:ngalM} and is
discussed further in Section~\ref{sec:repr-cosm-field}) for which we
are not able to determine the position of the peak.

In order to constrain the peak position for \textit{all} redshift
bins, and to provide an additional check on the robustness of our
results, we determine the peak position by using an alternative
abundance matching technique
\citep{Kravtsovetal04,Vale:2006fm,2006ApJ...647..201C}.  Essentially
one matches the abundances of haloes selected in a certain way to
galaxies selected in a (hopefully equivalent) way.

For our abundance matching analysis, we use a series of snapshot
outputs at each of our redshift slices from a large, high resolution
$N-$body dark-matter-only simulation performed with {\sc Gadget-2}
\citep{2005MNRAS.364.1105S} for a $\Lambda$CDM universe using Planck
parameters \citep{PlanckCollaboration:2014dt}, namely $\Omega_M=0.307$,
$\Omega_{\Lambda}=0.693$, $h=0.678$ and $\sigma_8=0.829$. The size of
the simulation box is a cube of 80 $h^{-1}$~Mpc on a side and contains
in total 1024$^3$ particles with a mass resolution of $3\times
10^7M_{\odot}$ per particle. Haloes and sub-haloes are identified
using the halo finding algorithm {\sc AdaptaHOP}
\citep{2004MNRAS.352..376A} which uses an SPH-like kernel to compute
densities at the location of each particle and partitions the ensemble
of particles into halos and sub-haloes based on saddle points in the
density field. The minimum number of particles per halo is 20:
therefore, the least massive haloes in our survey ($\sim~10^{11}
M_\odot$) are well resolved.

Next, circular velocities ($v_{\rm max}$) and masses ($M_{\rm 200}$)
are extracted for each halo and subhalo. Circular velocities are
defined as in the usual way, $V_{\rm max}=\rm max(\sqrt{Gm(\leq
  r)/r}\,)$, where $m(\leq r)$ is the mass enclosed at radius
r. $V_{\rm max}$ can be estimated without any accurate estimate of the
physical boundary of the objects which be difficult in particular for
sub-haloes. For each object, we define the radius $R_{200}$ (and thus
the mass $M_{200}$) as the radius where the enclosed mean density
$M_V/(4\pi R_V^3/3)$ is $200$ times the critical density,
$\rho_c(z)=3H(z)^2/8\pi G$, where
$H(z)=H_0\sqrt{\Omega_m(1+z)^3+\Omega_\Lambda}$.

At each redshift bin, we determine the stellar mass threshold which
matches the total abundance of galaxies selected by stellar mass $M$
to the total abundances of haloes selected by $v_{\rm max}$, i.e.,

\begin{equation}
N_h(>V_{\rm max})= N_{\rm g}(>M) . 
\end{equation}

We compute our galaxy abundances by integrating the mass functions
given in \cite{Ilbert:2013dq}.  Then, at each redshift slice, we fit a
simple linear function to the relationship between the median
$v_{\rm max}$ and $M_{200}$ for each bin of median halo mass. This
allows us in turn to derive the characteristic halo masses at each
abundance threshold, and, consequently, at the corresponding stellar
mass threshold. The results from this procedure are shown as the solid
lines in Figure~\ref{fig:shmr} (this line is actually the fit to the
\cite{2003MNRAS.339.1057Y} relation). At each redshift slice our
simulation contains sufficient numbers of low-mass and high-mass
haloes to reliably constrain the location the position of the peak in
the ratio $M_*/M_h$. The arrows on each panel shows the completeness
limits in stellar mass threshold presented in
Figure~\ref{fig:selectionVista}.

At $z<1$ abundance matching measurements agree with our halo model
measurements for higher-mass haloes: at the lower-mass end there is a
slight systematic offset. We note that the halo mass function used for
our halo modelling is not the same as the halo mass function in our
HOD model. As the dark matter halo mass function as these redshifts is
not constrained by observations, it is difficult to choose between
these two mass functions.  We note that in the high mass regime, the
two methods are in good agreement, suggesting that $M_{\rm min}$ and
$M_{\rm 200}$ are equivalent estimates of halo mass.

\begin{figure}
\begin{center}
  \includegraphics[width=0.49\textwidth]{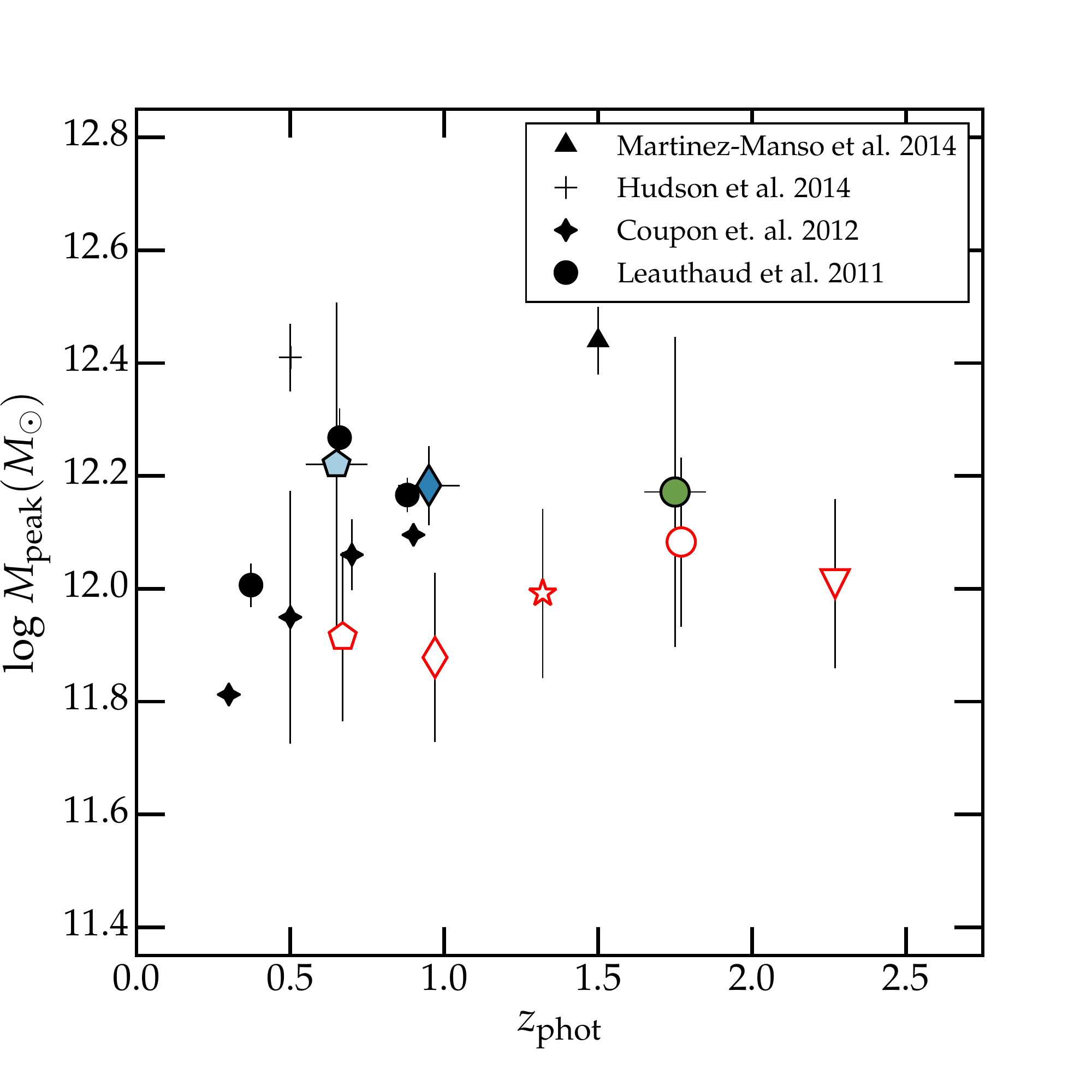}
\end{center}
\caption{Location of the maximum in $M^*/M_h$ from the HOD fitting
  procedure (filled coloured symbols) and for the abundance matching
  (open red symbols). Also shown are a selection of literature
  results. For clarity, the abundance matching measurements are
  slightly offset in redshift from the HOD points. }
\label{fig:shmrpeak}
\end{figure}

As before, each redshift bin, we fitted the position of the peak using
the \citeauthor{2003MNRAS.339.1057Y} analytic expression. These points
are shown as the open symbols in Figure~\ref{fig:shmrpeak}, slightly
offset for clarity. Figure~\ref{fig:shmrpeak} also includes a
selection of literature measurements.  We note the large scatter
between previous measurements, which is probably related either to the
measurement technique or the sample selection. Most lower-redshift
samples, with the exception of \citep{Leauthaud:2011fz}, are
luminosity-selected and not mass-selected, and the conversion to a
reliable mass-selected sample is uncertain (see Figure 14. in
\citeauthor{Couponetal12} for an idea of the typical
uncertainties). 

We should also note that in this work we compute the dark matter halo
masses given a sample of galaxies selected by stellar mass. Works such
as \cite{Leauthaudetal12} actually calculate stellar mass content for
a given halo mass. In the case of large scatter between stellar mass
and halo mass, these two measurements may not be equivalent. 

In this work, our measurements are always made for highly complete
samples. Stellar mass errors can potentially have an effect on the
derived SHMR. This effect has been treated in detail in
\cite{Behroozietal10}. The effect of stellar mass errors on derived
mass function in this present data set has been described in detail in
\cite{Ilbert:2013dq}. The most pernicious effect is the ``Eddington
Bias'' \citep{Eddington:1913tz} which can affect the high mass end of
the stellar mass function. Figure A2 in \citeauthor{Ilbert:2013dq}
shows that a simple gaussian description of stellar mass errors
$\sigma=0.04*(1.+z)$ results in at most a 0.1-0.2 dex overestimate in
stellar mass functions due to in only the most massive bins
($\logm\sim12$) -- and in these bins there are not sufficient numbers
of galaxies to measure correlation functions.

In summary, neither our HOD measurements nor our abundance matching
indicates an evolution in the position of $M_{\rm peak}$ as a function
of redshift, as have been claimed by previous authors (although, of
course, a small increase with redshift cannot be ruled out by our
measurements). The implication of this result will be discussed in
subsequent sections.

\subsection{The satellite fraction and its evolution with redshift}
\label{sec:satell-fract-its}

The physical properties of satellite galaxies (i.e., galaxies less
massive than the central galaxy but lying inside the same dark matter
halo) can provide additional information concerning the evolutionary
history of the host dark matter halo. It is now relatively well
established from both numerical simulations and observations that
physical processes can modify the number of satellite galaxies. Major
or minor mergers also play a role in affecting the satellite
fraction. Furthermore, it seems that at $z<1$, galaxy evolution is
mostly ``secular'', and major mergers may not be significant
\citep{LopezSanjuan:2011jf}. At these lower redshifts however
``environmental quenching'' processes
\citep{2005ApJ...625..621B,Peng:2010p11940} may modify the low-mass
end of the global mass function. However, the rapid evolution in the
global normalisation of the stellar mass functions between $1<z<2$
indicates that merging is an important process at higher redshifts,
and suggest that the satellite fractions in low-redshift and
high-redshift regimes should be different. From our halo occupation
distribution model, we can derive the fraction of galaxies in a given
dark matter halo which are satellite galaxies
(Equation~\ref{eq:frsat}). It still remains to be seen what is the
link between satellite fractions derived directly from
spectroscopically identified groups \cite{Kovac:2014iv} and
measurements made such as these: it is only possible to compare these
independent results to ours over a very limited range in halo mass.

\begin{figure}
\begin{center}
  \includegraphics[width=0.49\textwidth]{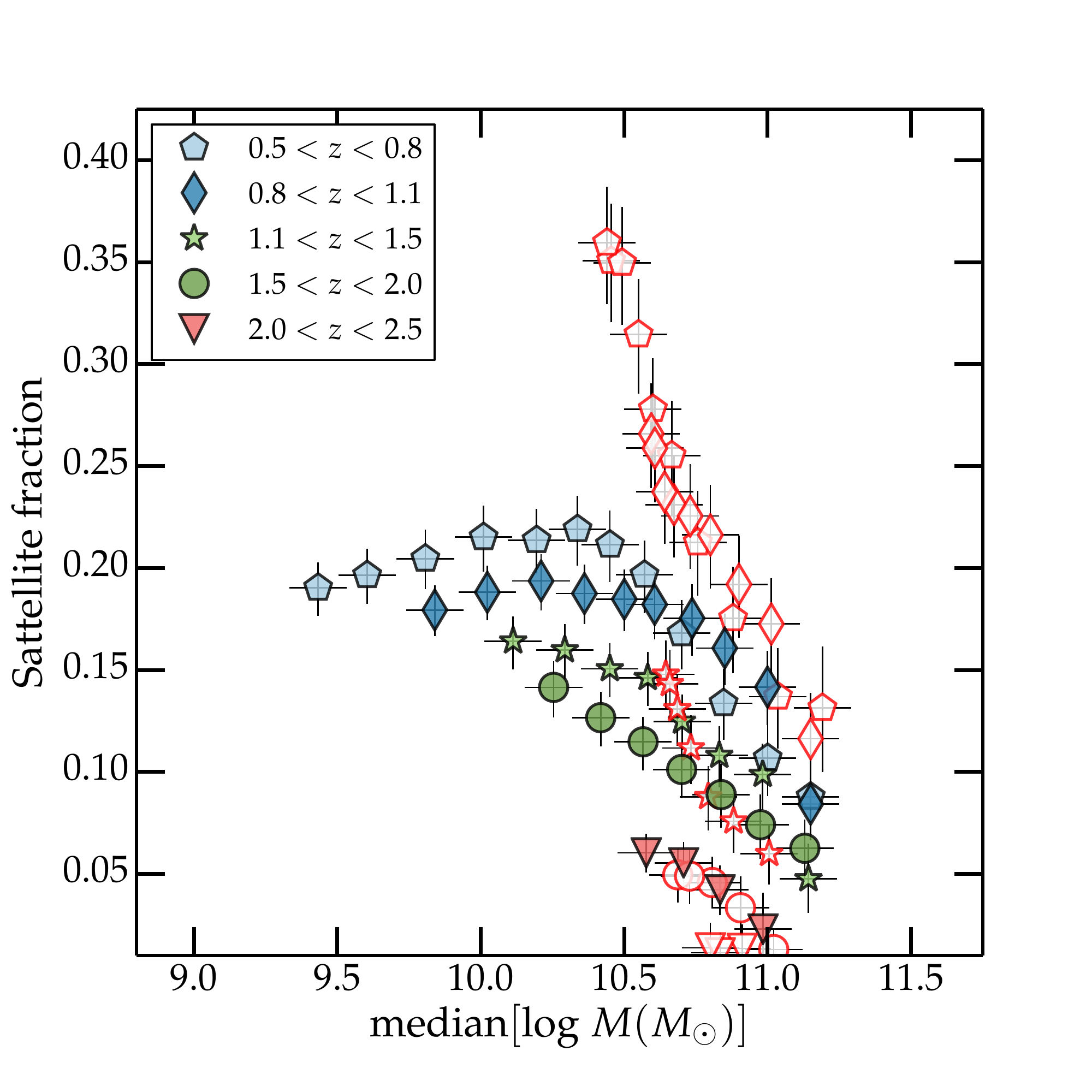}
\end{center}
\caption{The satellite fraction as a function of median stellar mass
  threshold for the full sample and for the passive galaxy sample
  (open symbols).}
\label{fig:frsat}
\end{figure}

\begin{figure}
\begin{center}
  \includegraphics[width=0.49\textwidth]{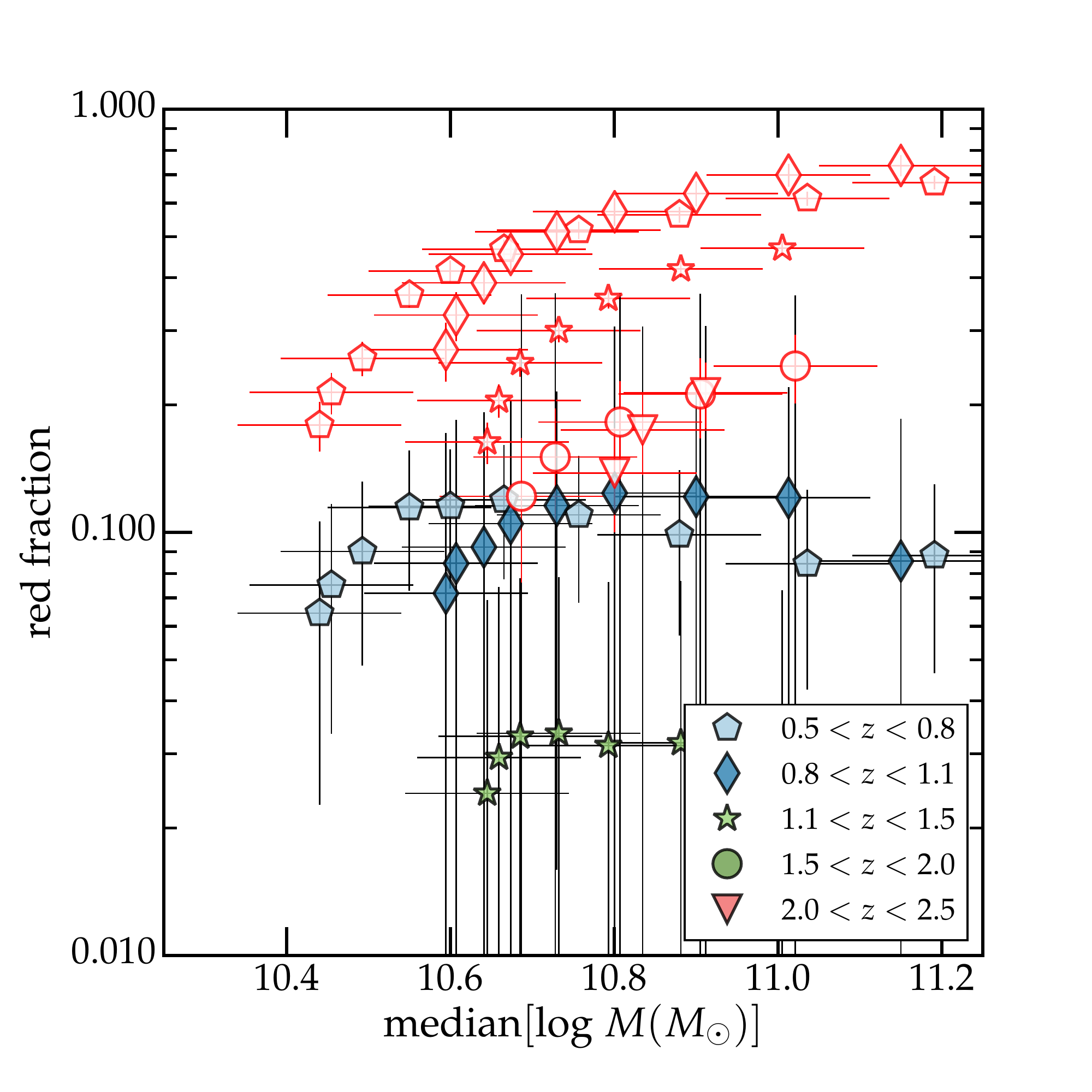}
\end{center}
\caption{The fraction of the total galaxy population which are passive
  (filled symbols) compared with the fraction of the total galaxy
  population which are passive satellites (open red symbols).}
\label{fig:frsat-passive}
\end{figure}

Figure~\ref{fig:frsat} shows the satellite fraction $f_{s}$ as a function of the
stellar mass for each redshift bin for both the passive galaxy sample and the
full sample. At all redshifts, the satellite fraction decreases as the stellar
mass threshold increases. We also note that at lower redshift bins, at a fixed
stellar mass threshold, the satellite fraction at intermediate stellar masses
($\logm\sim10.0-10.5$) is \textit{higher} in the quiescent sample than in the
full one. This trend is not seen at higher stellar masses, where the satellite
fraction remains low, regardless of the galaxy type. 

This trend is reversed above $z>1$, where the fraction of satellites is
\textit{lower} in quiescent galaxies populations than in the full sample. The
trends in satellite fractions seen here with selection by mass and
star-formation activity at low redshifts are broadly consistent with those seen
in \cite{Couponetal12} and \cite{2013ApJ...778...93T}, although the former study
made selections in ``corrected'' luminosity and rest-frame colour. Our
measurements are below those of \cite{Wakeetal11}; they fixed $\alpha$ in
Equation~\ref{eq:frsat} to one while in our case it is treated as a free
parameter. This flattening in the satellite fraction was also observed by
\cite{Wakeetal11}, \cite{Zehavietal11} and \cite{Zhengetal07}.

Since we know the fraction of satellite galaxies for both the full population
and the passive population, we can compute the fraction of the total galaxy
population which is a ``passive satellite'' and also (using the total number of
galaxies) the fraction of passive galaxies. Figure~\ref{fig:frsat-passive} shows
the fraction of the total galaxy population which is passive (open red symbols)
and the fraction of total galaxy population which are passive satellite galaxies
(filled symbols). The dependence of the passive fraction on mass is simply a
reflection of the well-known result that the peak in the number of quiescent
galaxies is at $z\sim0.8$. This is to some extent mirrored in the evolution of
the fraction of passive satellite galaxies, which tracks the overall passive
galaxy population. In all cases, the fraction of passive satellite galaxies
drops steeply at higher redshifts. Taken together, these trends suggest that
massive galaxies at high redshifts may have already accreted all their
satellites.


\subsection{Galaxy bias}

Galaxies are not perfect tracers of the underlying dark matter
distribution. (Depending on one's viewpoint, this may be regarded
either as a ``nuisance parameter'' or containing information
concerning galaxy evolution.)  A knowledge of galaxy bias has become
important in calibrating accurately cosmological probes, and so we now
turn to a determination of galaxy bias in our survey using the halo
model.

\begin{figure}
\begin{center}
  \includegraphics[width=0.49\textwidth]{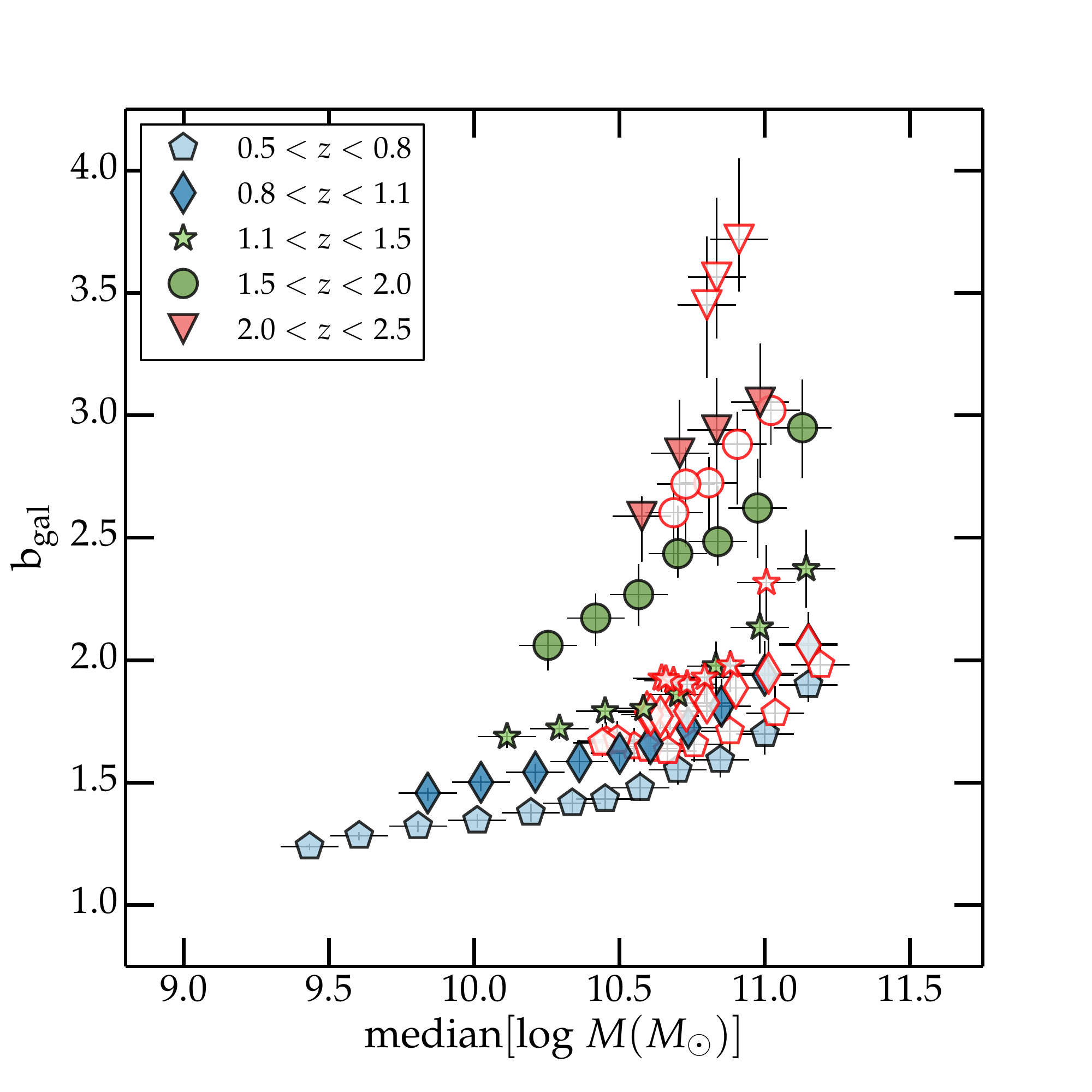}
\end{center}
\caption{The galaxy bias for each redshift slice for the full and
  quiescent samples (filled and red open symbols respectively) as a
  function of median stellar mass and for each redshift slice.}
\label{fig:bgalsm}
\end{figure}

The well-known dependence of galaxy bias on luminosity has been
studied extensively both in the local Universe and at higher redshifts
\citep{Norbergetal01,Zehavietal11,Polloetal06, Coiletal08,
  Meneuxetal09, Marullietal13}. Photometric surveys have also provided
important information at higher redshifts \citep{2008A&A...479..321M,
  Couponetal12}. The consensus from these studies is that that bias is
a weak function of luminosity for galaxies with $L<L_{\ast}$ (where
$L_{\ast}$ is the characteristic luminosity from the Schechter
function) and increases steeply for $L>L_{\ast}$. Interpreting these
bias measurements has not always been straightforward, because in
luminosity-selected surveys substantial luminosity evolution with
redshift complicates our understanding of the relationship between
mass in stars and galaxy mass. It is only very recently that it has
been possible to make bias measurements as a function of stellar mass
for a statistically significant volume.

Figure~\ref{fig:bgalsm} shows the galaxy bias derived from our
best-fitting halo model parameters (see Equation~\ref{eq:galbias}) as
a function of the stellar mass at each redshift bin.  We see a
monotonic trend that bias increases with redshift as in
\cite{ArnalteMur:2014db} for example. For $z \le 1$, and for stellar
masses less than $10^{10.7}M_{\sun}$ for the full sample and for
$10^{10.7}M_{\sun}$ for the quiescent sample, the bias depends weakly
on stellar mass. However at stellar masses $10^{10.7 -- 10.9}M_{\sun}$
bias is a strong function of stellar mass. For low stellar mass
threshold full galaxy samples selected at low redshift we find
$b_{gal} \sim 1.3$; the quiescent population is more strongly biased
$b_{gal}\sim 1.6$. in agreement with the CFHTLS and PRIMUS
\cite{Couponetal12,Skibbaetal13}.

Our data provides the most reliable measurement to date of the bias of
quiescent galaxy populations at high redshifts: at $2.0<z<2.5$ we find
$b_{\rm gal}\sim3.5$ for the quiescent galaxy population, and $b_{\rm
  gal}\sim3$ for the full galaxy population. We note also that the
difference in bias between the full galaxy population and the
quiescent population at the same stellar mass threshold increases at
higher redshifts. This is almost certainly because the full galaxy
population is dominated by star-forming galaxies at high redshifts,
which are weakly clustered, a point to which we will return to in the
discussion Section.

\subsection{On the representativeness of the COSMOS field and a
  comparison with other results}
\label{sec:repr-cosm-field}

It has already been noted that there are an overabundance of rich
structures in the cosmos field \citep{Meneuxetal09,hjmccetal07},
particularly at $z\sim1$. In fact, some earlier studies such as
\citeauthor{Meneuxetal09} failed to find any significant dependence of
clustering on stellar mass threshold, which in all likelihood due to
their relatively bright threshold in stellar mass, but also to the
overabundance of rich clusters in the field. More speculatively, the
discovery of a ``quasar wall'' a few degrees away from the COSMOS
field which may in part give rise to the overal increase in density in
the COSMOS field \citep{Clowes:2013jp}. However, despite this overall
slightly elevated density increase, most of the relations presented in
previous sections are qualitatively in agreement with what one expect,
with the exception of the redshift bin at $1<z<1.5$ which deviates
markedly and consistently from all trends presented in this paper.

To investigate further the discrepancy at $1<z<1.5$ we used the
``WIRDS'' data set \citep{2012A&A...545A..23B} to investigate the
nature of the differences between COSMOS and other data. Although
WIRDS is shallower than the present data set, it consists of four
fields separated widely on the sky. We measured correlation functions
in averaged mass-thresholded data in WIRDS and compared it to the
COSMOS field. Interestingly, we find that the WIRDS data agrees with
COSMOS on large scales, but a small scales there is \textit{much} more
power in COSMOS than in WIRDS. The implication which this has in the
halo model fits can be seen in Figure~\ref{fig:Mhngal}: for a given
abundance, the derived halo masses are much lower than one would
expect, essentially because of the much steeper one-halo term. This
has implications for any study using the COSMOS field to measure small
scale clustering at $1<z<1.5$. For example, the very steep two-point
correlation function measured in
\citep{2010ApJ...708..202M,Bethermin:2014dh} at $1<z<1.5$ seems now to
be an artefact of very rich, small scale clustering present at this
redshift bin.

\section{Discussion: the changing relationship between galaxies and
  the dark matter haloes they inhabit}

To understand the results presented in the previous Sections they need
to be considered in the general context of the evolution of the galaxy
population from $z\sim~0$ to $z\sim2$, some aspects of which are now
quite well understood. Of course, the precise \textit{mechanisms}
which give rise to these changes is still debated, but we can at least
consider how our abundance and halo mass measurements reflect these
well-established changes in the global galaxy population.

It is worth starting by reminding ourselves of the evolution observed
in the stellar mass function from $z\sim0$ to $z\sim2$ for both
passive and full galaxy populations. As described in
\cite{Ilbert:2013dq} (based on measurements made in these data), the
total normalisation of the global mass function increases steadily
from $z\sim2$ to $z\sim0$. However, for the passive galaxy population
a rapid build-up is observed in the faint-end slope below $z\sim1$. We
may also consider the evolution of the amount of star-formation per
unit mass (sSFR), the specific star formation rate, sSFR. At a given
stellar mass, the sSFR declines steadily until $z\sim0$
\citep{Ilbert:2013dq}: stated another way, star formation at high
redshift occurs preferentially in higher-mass systems. The
implications for the results presented here are several: firstly, at a
fixed stellar mass threshold, the proportions of passive and
star-forming galaxies is a strong function of redshift and stellar
mass. At high redshifts ($z\sim2$), our mass-selected samples are
dominated by star-forming massive galaxies; at low redshifts, in
contrast, the low-mass end becomes increasingly dominated by
passively-evolving galaxies. How may we understand our results in
terms of these changes in the global galaxy population?

Firstly, it is interesting to consider the dependence of clustering
amplitude on stellar mass and redshift. Our measurements show clearly
the dependence of galaxy clustering on stellar mass threshold. It is
interesting to note that the weak dependence of clustering strength on
stellar mass for the passive galaxy population in our threshold
samples at $z\sim0.8$ is entirely due to the large ``lump'' of
passively-evolving galaxies at $M\sim 10^{10.5}$ $M_{\sun}$ visible in
the mass-redshift plane (Figure~\ref{fig:selectionVista}): changes in
the threshold do not change the number of lower mass galaxies in the
sample. Over the large redshift range of our survey, at a fixed
stellar mass, the projected clustering amplitude at a fixed stellar
mass threshold drops significantly. This is in due part a simple
projection effect (and note also that at higher redshifts, our
redshifts bins are larger) as our co-moving correlation lengths at
fixed stellar mass, measured using our halo model, remain remarkably
constant (Figure~\ref{fig:r0}). The increasingly biased nature of the
galaxy population, (Figure~\ref{fig:bgalsm}) is almost perfectly
offset by the decreasing clustering amplitude of the underlying dark
matter.

We fit our observations to a phenomenological halo model. Previously,
\citeauthor{Couponetal12} found that the fitted parameters of this
model changed remarkably little from $z\sim1$ to $z\sim0$, reflecting
the nature of the changes taking place in the galaxy population over
this redshift range. This is certainly not the these data: above
$z>1$, many aspects of the fitted model parameters change
radically. At fixed abundance, characteristic halo masses drop
significantly above $z>1$ (Figure~\ref{fig:Mhngal}). The mass fraction
of satellite galaxies contained in haloes drops almost to zero, and
this effect is ever more pronounced for the passive galaxy
population. In addition, the fraction of mass in satellite galaxies
for faint passively-evolving galaxies rises rapidly for faint galaxies
(one sees almost exactly the same effect if one considers these
quantities as a function of halo mass). It is challenging to compare
the satellite fractions measured in previous works
\citep{Kovac:2014iv} because of the different mass ranges probed with
respect to spectroscopic surveys. However, qualitatively, this
behaviour is in agreement with what one would expect from our current
hierarchical models of galaxy evolution, where at high redshifts dark
matter haloes are dominated by single massive galaxies. The rapid
increase in the satellite fraction for faint passive galaxies happens
exactly at the redshift range where the faint end of the mass function
rises sharply and ``satellite quenching'' processes become a dominant
process in galaxy evolution \citep{Peng:2010p11940}.

Considering the differences between the passive and total galaxy
population, it is interesting to note that passive galaxies and the
full galaxy population lie on the same halo mass / abundance
relationship (Figure~\ref{fig:Mhngal}), and occupy the same region of
parameter space as ``normal'' galaxies. In almost all of the plots
presented in this paper, the passive galaxy sample occupies the same
region of parameter space as a less abundant, more clustered version
of the full galaxy sample with one important exception: the satellite
fraction, where the passive population is revealed as radically
different from the full galaxy population. 

We also consider the ratio between stellar mass and halo mass for a
range of halo masses in each of our redshift slices. We find that the
peak position shows only a weak dependence on redshift. This is a
natural consequence of the fact that the shape of the overal stellar
mass function and halo mass function evolves little from $z\sim2$ to
the present day. We compared our halo model measurements with the
results of an abundance matching technique and find approximately the
same behaviour. The abundance matching measurements allow us estimate
$M^*/M_h$ ratios even for redshifts bins for which we are not able to
fit our halo model results. In general, results are broadly in
agreement with models which attempt to model jointly with the
evolution of the stellar mass function over large redshift
baseline. For example \citep{Behroozietal13} find only a small
increase in characteristic halo mass with redshift.

\section{Conclusions}

We have used highly precise photometric redshifts in the UltraVISTA-DR1
near-infrared survey to investigate the changing relationship between
galaxy stellar mass and the dark matter haloes hosting them to
$z\sim2$. We have achieved this by measuring the clustering properties
and abundances of a series of volume-limited galaxy samples selected
by stellar mass and star-formation activity. These measurements span a
uniquely large range in stellar mass and redshift and reach below the
characteristic stellar mass to $z\sim2$.

We found the following results: 1. At fixed redshift and scale,
clustering amplitude depends monotonically on sample stellar mass
threshold; 2. At fixed angular scale, the projected clustering
amplitude decreases with redshift but the co-moving correlation length
remains constant; 3. Characteristic halo masses and galaxy bias
increase with increasing median stellar mass of the sample; 4. The
slope of these relationships is modified in lower mass haloes;
5. Concerning the passive galaxy population, characteristic halo
masses are consistent with a simply less-abundant version of the full
galaxy sample, but at lower redshifts the fraction of satellite
galaxies in the passive population is very different from the full
galaxy sample; 6. Finally we find that the ratio between the
characteristic halo mass and median stellar mass at each redshift bin
reaches a peak at ${\log(M_h/M_\odot)}\sim12.2$ and the position of
this peak remains constant out to $z\sim2$. The behaviour of the full
and passively evolving galaxy samples can be understood qualitatively
by considering the slow evolution of the characteristic stellar mass
in the redshift range covered by our survey.

The next step is to extend this analysis to higher redshifts ($z>4$),
where the discrepancy between models of galaxy formation and
observations becomes even more acute. The new UltraVISTA DR2 data
release, reaching several magnitudes deeper in all near-infrared
bands, will enable for the first time this kind of study.

\section{Acknowledgments}

HJM acknowledges financial support form the ``Programme national
cosmologie et galaxies'' (PNCG). This work is based on data products
from observations made with ESO Telescopes at the La Silla Paranal
Observatory under ESO programme ID 179.A-2005 and on data products
produced by TERAPIX and the Cambridge Astronomy Survey Unit on behalf
of the UltraVISTA consortium. JSD acknowledges the support of the
European Research Council via the award of an Advanced Grant, and the
contribution of the EC FP7 SPACE project ASTRODEEP (Ref.No: 312725).

\bibliographystyle{mn2e}

\begin{thebibliography}{}

\bibitem[\protect\citeauthoryear{Arnalte-Mur et~al.,}{Arnalte-Mur
  et~al.}{2014}]{ArnalteMur:2014db}
Arnalte-Mur P.  et~al., 2014, \mnras, 441, 1783

\bibitem[\protect\citeauthoryear{Aubert, Pichon \& Colombi}{Aubert
  et~al.}{2004}]{2004MNRAS.352..376A}
Aubert D.,  Pichon C.,    Colombi S.,  2004, \mnras, 352, 376

\bibitem[\protect\citeauthoryear{{Behroozi}, {Conroy} \& {Wechsler}}{{Behroozi}
  et~al.}{2010}]{Behroozietal10}
{Behroozi} P.~S.,  {Conroy} C.,    {Wechsler} R.~H.,  2010, \apj, 717, 379

\bibitem[\protect\citeauthoryear{{Behroozi}, {Wechsler} \& {Conroy}}{{Behroozi}
  et~al.}{2013}]{Behroozietal13}
{Behroozi} P.~S.,  {Wechsler} R.~H.,    {Conroy} C.,  2013, \apj, 770, 57

\bibitem[\protect\citeauthoryear{Benjamin, Van~Waerbeke, M{\'e}nard \&
  Kilbinger}{Benjamin et~al.}{2010}]{2010MNRAS.408.1168B}
Benjamin J.,  Van~Waerbeke L.,  M{\'e}nard B.,    Kilbinger M.,  2010, \mnras,
  408, 1168

\bibitem[\protect\citeauthoryear{B{\'e}thermin et~al.,}{B{\'e}thermin
  et~al.}{2014}]{Bethermin:2014dh}
B{\'e}thermin M.  et~al., 2014, \aap, 567, A103

\bibitem[\protect\citeauthoryear{Bielby et~al.,}{Bielby
  et~al.}{2012}]{2012A&A...545A..23B}
Bielby R.  et~al., 2012, \aap, 545, 23

\bibitem[\protect\citeauthoryear{Bielby et~al.,}{Bielby
  et~al.}{2014}]{Bielby:2014dv}
Bielby R.~M.  et~al., 2014, \aap, 568, A24

\bibitem[\protect\citeauthoryear{{Bruzual} \& {Charlot}}{{Bruzual} \&
  {Charlot}}{2003}]{Bruzualetal03}
{Bruzual} G.,  {Charlot} S.,  2003, \mnras, 344, 1000

\bibitem[\protect\citeauthoryear{Bruzual \& Charlot}{Bruzual \&
  Charlot}{2003}]{Bruzual:2003p963}
Bruzual G.,  Charlot S.,  2003, \mnras, 344, 1000

\bibitem[\protect\citeauthoryear{Bundy, Ellis \& Conselice}{Bundy
  et~al.}{2005}]{2005ApJ...625..621B}
Bundy K.,  Ellis R.~S.,    Conselice C.~J.,  2005, \apj, 625, 621

\bibitem[\protect\citeauthoryear{Capak et~al.,}{Capak
  et~al.}{2007}]{2007ApJS..172...99C}
Capak P.  et~al., 2007, \apjs, 172, 99

\bibitem[\protect\citeauthoryear{Chabrier}{Chabrier}{2003}]{Chabrier:2003ki}
Chabrier G.,  2003, \pasp, 115, 763

\bibitem[\protect\citeauthoryear{Clowes, Harris, Raghunathan, Campusano,
  Sochting \& Graham}{Clowes et~al.}{2013}]{Clowes:2013jp}
Clowes R.~G.,  Harris K.~A.,  Raghunathan S.,  Campusano L.~E.,  Sochting
  I.~K.,    Graham M.~J.,  2013, \mnras, 429, 2910

\bibitem[\protect\citeauthoryear{{Coil} et~al.,}{{Coil}
  et~al.}{2008}]{Coiletal08}
{Coil} A.~L.  et~al., 2008, \apj, 672, 153

\bibitem[\protect\citeauthoryear{Conroy \& Wechsler}{Conroy \&
  Wechsler}{2009}]{2009ApJ...696..620C}
Conroy C.,  Wechsler R.~H.,  2009, \apj, 696, 620

\bibitem[\protect\citeauthoryear{Conroy, Wechsler \& Kravtsov}{Conroy
  et~al.}{2006}]{2006ApJ...647..201C}
Conroy C.,  Wechsler R.~H.,    Kravtsov A.~V.,  2006, \apj, 647, 201

\bibitem[\protect\citeauthoryear{{Cooray} \& {Sheth}}{{Cooray} \&
  {Sheth}}{2002}]{Coorayetal02}
{Cooray} A.,  {Sheth} R.,  2002, \physrep, 372, 1

\bibitem[\protect\citeauthoryear{{Coupon} et~al.,}{{Coupon}
  et~al.}{2009}]{Couponetal09}
{Coupon} J.  et~al., 2009, \aap, 500, 981

\bibitem[\protect\citeauthoryear{{Coupon} et~al.,}{{Coupon}
  et~al.}{2012}]{Couponetal12}
{Coupon} J.  et~al., 2012, \aap, 542, A5

\bibitem[\protect\citeauthoryear{Cowie, Songaila, Hu \& Cohen}{Cowie
  et~al.}{1996}]{Cowie:1996p8471}
Cowie L.~L.,  Songaila A.,  Hu E.~M.,    Cohen J.~G.,  1996, Astronomical
  Journal v.112, 112, 839

\bibitem[\protect\citeauthoryear{Croton, Gao \& White}{Croton
  et~al.}{2007}]{Croton:2007p11858}
Croton D.~J.,  Gao L.,    White S. D.~M.,  2007, \mnras, 374, 1303

\bibitem[\protect\citeauthoryear{Daddi, Cimatti, Renzini, Fontana, Mignoli,
  Pozzetti, Tozzi \& Zamorani}{Daddi et~al.}{2004}]{Daddi:2004p76}
Daddi E.,  Cimatti A.,  Renzini A.,  Fontana A.,  Mignoli M.,  Pozzetti L.,
  Tozzi P.,    Zamorani G.,  2004, \apj, 617, 746

\bibitem[\protect\citeauthoryear{Davis et~al.,}{Davis
  et~al.}{2003}]{2003SPIE.4834..161D}
Davis M.  et~al., 2003, Discoveries and Research Prospects from 6- to
  10-Meter-Class Telescopes II. Edited by Guhathakurta, 4834, 161

\bibitem[\protect\citeauthoryear{Eddington}{Eddington}{1913}]{Eddington:1913tz}
Eddington A.~S.,  1913, \mnras

\bibitem[\protect\citeauthoryear{Foucaud, Conselice, Hartley, Lane, Bamford,
  Almaini \& Bundy}{Foucaud et~al.}{2010}]{Foucaud:2010p11288}
Foucaud S.,  Conselice C.~J.,  Hartley W.~G.,  Lane K.~P.,  Bamford S.~P.,
  Almaini O.,    Bundy K.,  2010, \mnras, 406, 147

\bibitem[\protect\citeauthoryear{{Groth} \& {Peebles}}{{Groth} \&
  {Peebles}}{1977}]{GP77}
{Groth} E.~J.,  {Peebles} P.~J.~E.,  1977, ApJ, 217, 385

\bibitem[\protect\citeauthoryear{Guzzo et~al.,}{Guzzo
  et~al.}{2014}]{Guzzo:2014eb}
Guzzo L.  et~al., 2014, \aap, 566, A108

\bibitem[\protect\citeauthoryear{{Hoaglin}, {Mosteller} \& {Tukey}}{{Hoaglin}
  et~al.}{1983}]{Hoaglinetal83}
{Hoaglin} D.~C.,  {Mosteller} F.,    {Tukey} J.~W.,  1983, {Understanding
  robust and exploratory data anlysis}

\bibitem[\protect\citeauthoryear{{Ilbert} et~al.,}{{Ilbert}
  et~al.}{2006}]{Ilbertetal06}
{Ilbert} O.  et~al., 2006, \aap, 457, 841

\bibitem[\protect\citeauthoryear{Ilbert et~al.,}{Ilbert
  et~al.}{2013}]{Ilbert:2013dq}
Ilbert O.  et~al., 2013, \aap, 556, A55

\bibitem[\protect\citeauthoryear{Ilbert et~al.,}{Ilbert
  et~al.}{2010}]{2010ApJ...709..644I}
Ilbert O.  et~al., 2010, \apj, 709, 644

\bibitem[\protect\citeauthoryear{{Kilbinger} et~al.,}{{Kilbinger}
  et~al.}{2010}]{Kilbingeretal10}
{Kilbinger} M.  et~al., 2010, \mnras, 405, 2381

\bibitem[\protect\citeauthoryear{Kovac et~al.,}{Kovac
  et~al.}{2014}]{Kovac:2014iv}
Kovac K.  et~al., 2014, \mnras, 438, 717

\bibitem[\protect\citeauthoryear{{Kravtsov}, {Berlind}, {Wechsler}, {Klypin},
  {Gottl{\"o}ber}, {Allgood} \& {Primack}}{{Kravtsov}
  et~al.}{2004}]{Kravtsovetal04}
{Kravtsov} A.~V.,  {Berlind} A.~A.,  {Wechsler} R.~H.,  {Klypin} A.~A.,
  {Gottl{\"o}ber} S.,  {Allgood} B.,    {Primack} J.~R.,  2004, \apj, 609, 35

\bibitem[\protect\citeauthoryear{{Landy} \& {Szalay}}{{Landy} \&
  {Szalay}}{1993}]{Landyetal93}
{Landy} S.~D.,  {Szalay} A.~S.,  1993, ApJ, 412, 64

\bibitem[\protect\citeauthoryear{{Le F{\`e}vre} et~al.,}{{Le F{\`e}vre}
  et~al.}{2005}]{Lefevretal05b}
{Le F{\`e}vre} O.  et~al., 2005, \aap, 439, 845

\bibitem[\protect\citeauthoryear{{Leauthaud} et~al.,}{{Leauthaud}
  et~al.}{2010}]{Leauthaudetal10}
{Leauthaud} A.  et~al., 2010, \apj, 709, 97

\bibitem[\protect\citeauthoryear{Leauthaud, Tinker, Behroozi, Busha \&
  Wechsler}{Leauthaud et~al.}{2011}]{Leauthaud:2011bj}
Leauthaud A.,  Tinker J.,  Behroozi P.~S.,  Busha M.~T.,    Wechsler R.~H.,
  2011, \apj, 738, 45

\bibitem[\protect\citeauthoryear{Leauthaud et~al.,}{Leauthaud
  et~al.}{2011}]{Leauthaud:2011fz}
Leauthaud A.  et~al., 2011, \apj, 744, 159

\bibitem[\protect\citeauthoryear{{Leauthaud} et~al.,}{{Leauthaud}
  et~al.}{2012}]{Leauthaudetal12}
{Leauthaud} A.  et~al., 2012, \apj, 744, 159

\bibitem[\protect\citeauthoryear{Limber}{Limber}{1954}]{Limber:1954p11344}
Limber D.~N.,  1954, \apj, 119, 655

\bibitem[\protect\citeauthoryear{L{\'o}pez-Sanjuan et~al.,}{L{\'o}pez-Sanjuan
  et~al.}{2011}]{LopezSanjuan:2011jf}
L{\'o}pez-Sanjuan C.  et~al., 2011, \aap, 530, A20

\bibitem[\protect\citeauthoryear{McCracken et~al.,}{McCracken
  et~al.}{2010}]{2010ApJ...708..202M}
McCracken H.~J.  et~al., 2010, \apj, 708, 202

\bibitem[\protect\citeauthoryear{McCracken, Ilbert, Mellier, Bertin, Guzzo,
  Arnouts, Le~F{\`e}vre \& Zamorani}{McCracken
  et~al.}{2008}]{2008A&A...479..321M}
McCracken H.~J.,  Ilbert O.,  Mellier Y.,  Bertin E.,  Guzzo L.,  Arnouts S.,
  Le~F{\`e}vre O.,    Zamorani G.,  2008, \aap, 479, 321

\bibitem[\protect\citeauthoryear{McCracken et~al.,}{McCracken
  et~al.}{2012}]{McCracken:2012gd}
McCracken H.~J.  et~al., 2012, \aap, 544, A156

\bibitem[\protect\citeauthoryear{{Marulli} et~al.,}{{Marulli}
  et~al.}{2013}]{Marullietal13}
{Marulli} F.  et~al., 2013, \aap, 557, A17

\bibitem[\protect\citeauthoryear{{McCracken} et~al.,}{{McCracken}
  et~al.}{2007}]{hjmccetal07}
{McCracken} H.~J.  et~al., 2007, \apjs, 172, 314

\bibitem[\protect\citeauthoryear{{Meneux} et~al.,}{{Meneux}
  et~al.}{2009}]{Meneuxetal09}
{Meneux} B.  et~al., 2009, \aap, 505, 463

\bibitem[\protect\citeauthoryear{{Moster}, {Somerville}, {Maulbetsch}, {van den
  Bosch}, {Macci{\`o}}, {Naab} \& {Oser}}{{Moster} et~al.}{2010}]{Mosteretal10}
{Moster} B.~P.,  {Somerville} R.~S.,  {Maulbetsch} C.,  {van den Bosch} F.~C.,
  {Macci{\`o}} A.~V.,  {Naab} T.,    {Oser} L.,  2010, \apj, 710, 903

\bibitem[\protect\citeauthoryear{{Navarro}, {Frenk} \& {White}}{{Navarro}
  et~al.}{1997}]{Navarroetal97}
{Navarro} J.~F.,  {Frenk} C.~S.,    {White} S.~D.~M.,  1997, \apj, 490, 493

\bibitem[\protect\citeauthoryear{Neyman \& Scott}{Neyman \&
  Scott}{1952}]{Neyman:1952p11742}
Neyman J.,  Scott E.~L.,  1952, \apj, 116, 144

\bibitem[\protect\citeauthoryear{Norberg et~al.,}{Norberg
  et~al.}{2002}]{Norberg:2002p402}
Norberg P.  et~al., 2002, \mnras, 332, 827

\bibitem[\protect\citeauthoryear{{Norberg} et~al.,}{{Norberg}
  et~al.}{2001}]{Norbergetal01}
{Norberg} P.  et~al., 2001, \mnras, 328, 64

\bibitem[\protect\citeauthoryear{{Norberg}, {Gazta{\~n}aga}, {Baugh} \&
  {Croton}}{{Norberg} et~al.}{2011}]{Norbergetal11}
{Norberg} P.,  {Gazta{\~n}aga} E.,  {Baugh} C.~M.,    {Croton} D.~J.,  2011,
  \mnras, 418, 2435

\bibitem[\protect\citeauthoryear{Oke}{Oke}{1974}]{Oke:1974p12716}
Oke J.~B.,  1974, \apjs, 27, 21

\bibitem[\protect\citeauthoryear{Peacock \& Smith}{Peacock \&
  Smith}{2000}]{Peacock:2000p11176}
Peacock J.~A.,  Smith R.~E.,  2000, \mnras, 318, 1144

\bibitem[\protect\citeauthoryear{Peng et~al.,}{Peng
  et~al.}{2010}]{Peng:2010p11940}
Peng Y.-j.  et~al., 2010, \apj, 721, 193

\bibitem[\protect\citeauthoryear{{Planck Collaboration} et~al.,}{{Planck
  Collaboration} et~al.}{2014}]{PlanckCollaboration:2014dt}
{Planck Collaboration} et~al., 2014, \aap, 571, A16

\bibitem[\protect\citeauthoryear{{Polletta} et~al.,}{{Polletta}
  et~al.}{2007}]{Pollettaetal07}
{Polletta} M.  et~al., 2007, \apj, 663, 81

\bibitem[\protect\citeauthoryear{{Pollo} et~al.,}{{Pollo}
  et~al.}{2006}]{Polloetal06}
{Pollo} A.  et~al., 2006, \aap, 451, 409

\bibitem[\protect\citeauthoryear{{Roche}, {Eales}, {Hippelein} \&
  {Willott}}{{Roche} et~al.}{1999}]{Rocheetal99}
{Roche} N.,  {Eales} S.~A.,  {Hippelein} H.,    {Willott} C.~J.,  1999, \mnras,
  306, 538

\bibitem[\protect\citeauthoryear{Scoccimarro, Sheth, Hui \& Jain}{Scoccimarro
  et~al.}{2001}]{Scoccimarro:2001p11099}
Scoccimarro R.,  Sheth R.~K.,  Hui L.,    Jain B.,  2001, \apj, 546, 20

\bibitem[\protect\citeauthoryear{Scoville et~al.,}{Scoville
  et~al.}{2007}]{Scoville:2007p12720}
Scoville N.  et~al., 2007, \apjs, 172, 1

\bibitem[\protect\citeauthoryear{Seljak}{Seljak}{2000}]{Seljak:2000p1153}
Seljak U.,  2000, \mnras, 318, 203

\bibitem[\protect\citeauthoryear{{Sheth} \& {Tormen}}{{Sheth} \&
  {Tormen}}{1999}]{Sethetal99}
{Sheth} R.~K.,  {Tormen} G.,  1999, \mnras, 308, 119

\bibitem[\protect\citeauthoryear{{Skibba} et~al.,}{{Skibba}
  et~al.}{2013}]{Skibbaetal13}
{Skibba} R.~A.  et~al., 2013, ArXiv e-prints

\bibitem[\protect\citeauthoryear{Springel}{Springel}{2005}]{2005MNRAS.364.1105S}
Springel V.,  2005, \mnras, 364, 1105

\bibitem[\protect\citeauthoryear{Szalay, Connolly \& Szokoly}{Szalay
  et~al.}{1999}]{Szalay:1999p4804}
Szalay A.~S.,  Connolly A.~J.,    Szokoly G.~P.,  1999, \aj, 117, 68

\bibitem[\protect\citeauthoryear{Tinker, Leauthaud, Bundy, George, Behroozi,
  Massey, Rhodes \& Wechsler}{Tinker et~al.}{2013}]{2013ApJ...778...93T}
Tinker J.~L.,  Leauthaud A.,  Bundy K.,  George M.~R.,  Behroozi P.,  Massey
  R.,  Rhodes J.,    Wechsler R.~H.,  2013, \apj, 778, 93

\bibitem[\protect\citeauthoryear{{Tinker}, {Weinberg}, {Zheng} \&
  {Zehavi}}{{Tinker} et~al.}{2005}]{Tinkeretal05}
{Tinker} J.~L.,  {Weinberg} D.~H.,  {Zheng} Z.,    {Zehavi} I.,  2005, \apj,
  631, 41

\bibitem[\protect\citeauthoryear{Vale \& Ostriker}{Vale \&
  Ostriker}{2006}]{Vale:2006fm}
Vale A.,  Ostriker J.~P.,  2006, \mnras, 371, 1173

\bibitem[\protect\citeauthoryear{{Wake} et~al.,}{{Wake}
  et~al.}{2011}]{Wakeetal11}
{Wake} D.~A.  et~al., 2011, \apj, 728, 46

\bibitem[\protect\citeauthoryear{{Wraith}, {Kilbinger}, {Benabed}, {Capp{\'e}},
  {Cardoso}, {Fort}, {Prunet} \& {Robert}}{{Wraith}
  et~al.}{2009}]{Wraithetal09}
{Wraith} D.,  {Kilbinger} M.,  {Benabed} K.,  {Capp{\'e}} O.,  {Cardoso} J.-F.,
   {Fort} G.,  {Prunet} S.,    {Robert} C.~P.,  2009, \prd, 80, 023507

\bibitem[\protect\citeauthoryear{Yang, Mo \& van~den Bosch}{Yang
  et~al.}{2003}]{2003MNRAS.339.1057Y}
Yang X.,  Mo H.~J.,    van~den Bosch F.~C.,  2003, \mnras, 339, 1057

\bibitem[\protect\citeauthoryear{{Zehavi} et~al.,}{{Zehavi}
  et~al.}{2011}]{Zehavietal11}
{Zehavi} I.  et~al., 2011, \apj, 736, 59

\bibitem[\protect\citeauthoryear{Zentner, Hearin \& van~den Bosch}{Zentner
  et~al.}{2014}]{Zentner:2014ki}
Zentner A.~R.,  Hearin A.~P.,    van~den Bosch F.~C.,  2014, \mnras, 443, 3044

\bibitem[\protect\citeauthoryear{{Zheng} et~al.,}{{Zheng}
  et~al.}{2005}]{Zhengetal05}
{Zheng} Z.  et~al., 2005, \apj, 633, 791

\bibitem[\protect\citeauthoryear{{Zheng}, {Coil} \& {Zehavi}}{{Zheng}
  et~al.}{2007}]{Zhengetal07}
{Zheng} Z.,  {Coil} A.~L.,    {Zehavi} I.,  2007, \apj, 667, 760

\end{thebibliography}

{\noindent\small\it\ignorespaces $^1$ Institut d'Astrophysique de Paris, Universit\'e Pierre et
  Marie Curie - Paris 6, 98 bis Boulevard Arago, F-75014 Paris, France 
$^2$Institute for Astronomy, University of Hawaii, 2680 Woodlawn Drive, Honolulu, HI, 96822\\
$^3$CEA Saclay, Service d'Astrophysique (SAp), Orme des Merisiers,
 B\^{a}t. 709, 91191 Gif-sur-Yvette, France \\
$^4$Laboratoire d'Astrophysique de Marseille, UMR 7326, 
 38 rue Fr\'{e}d\'{e}ric Joliot-Curie, 13388 Marseille cedex 13,
 France\\
$^5$ Astronomical Observatory of the University of Geneva, ch. d'Ecogia  16, 1290 Versoix, Switzerland\\ 
$^6$ SUPA, Institute for Astronomy, University of Edinburgh,
  Royal Observatory, Edinburgh EH9 3HJL, UK\\
$^7$ Dark Cosmology Centre, Niels Bohr Institute,
  University of Copenhagen, Juliane Maries Vej 30, 2100 Copenhagen,
  Denmark\\
$^8$ Kapteyn Astronomical Institute, University of Groningen, P.O. Box 800, 9700
AV Groningen, The Netherlands\\
$^9$ European Southern Observatory, Karl-Schwarzschild-Str. 2, 85748 Garching, Germany\\
}
\appendix
\end{document}